\documentclass[%
 aip,
 reprint,
]{revtex4-2}

\usepackage{graphicx}
\usepackage{dcolumn}
\usepackage{mathtools}
\usepackage{amsmath,amssymb,amsbsy}
\usepackage{bbm,bm,mathrsfs,yfonts}
\usepackage[english]{babel}
\usepackage{hyperref}
\usepackage[capitalize]{cleveref}
\usepackage{xcolor}
\usepackage{microtype}
\newcommand{\bv}[1]{\mathbf{#1}}
\newcommand{\hv}[1]{\hat{\mathbf{#1}}}

\newcommand{\pol}[1]{\mathrm{#1}}
\newcommand{\mr}[1]{\mathrm{#1}}

\DeclarePairedDelimiter\floor{\lfloor}{\rfloor}
\usepackage[utf8]{inputenc}
\usepackage[T1]{fontenc}
\usepackage{mathptmx}

\begin{document}

\title{Four-dimensional drift-kinetic model for scrape-off layer plasmas}

\author{L. M. Perrone}
 \altaffiliation[Currently at ]{DAMTP, University of Cambridge, CMS, Wilberforce Road, Cambridge CB3 0WA, UK}
 \email{lmp61@cam.ac.uk}
\author{R. Jorge}
 \altaffiliation[Currently at ]{Institute for Research in Electronics and Applied Physics, University of Maryland, College Park, MD 20742, USA}
\author{P. Ricci}
\affiliation{ 
École Polytechnique Fédérale de Lausanne (EPFL), Swiss Plasma Center (SPC), CH-1015
Lausanne, Switzerland
}%

\begin{abstract}
A four-dimensional plasma model able to describe the scrape-off layer region of tokamak devices at arbitrary collisionality is derived in the drift-reduced limit. The basis of the model is provided by a drift-kinetic equation that retains the full non-linear Coulomb collision operator and describes arbitrarily far from equilibrium distribution functions. By expanding the dependence of distribution function over the perpendicular velocity in a Laguerre polynomial basis and integrating  over the perpendicular velocity, a set of four-dimensional moment equations for the expansion coefficients of the distribution function is obtained. The Coulomb collision operator, as well as Poisson's equation, are evaluated explicitly in terms of perpendicular velocity moments of the distribution function. 
\end{abstract}

\maketitle

\section{Introduction}

Understanding the plasma dynamics in the scrape-off layer (SOL), the most external plasma region in magnetic confinement devices, is of primary importance on the way to fusion energy.
In fact, this region plays an essential role in the overall performance of a fusion device by controlling the interaction of the plasma with the wall,  therefore regulating, among others, the impurity dynamics, the heat flux to the vessel walls, the fuelling, and the recycling process \cite{Stangeby2000}.
Improving our understanding of this region is considered as a crucial step on the way to fusion energy \cite{Ricci2015}. 
 With respect to the core plasma, the SOL is characterised by large amplitude fluctuations, including coherent filamentary structures, called blobs \cite{Theiler2011}, that develop on large spatial scales comparable to the time-averaged SOL pressure gradient length $L_p$ and on time scales below the ion cyclotron frequency, $\Omega_{ci}=e B/m_i$, being $e$, $B$, and $m_i$, the electron charge, magnetic field, and ion mass, respectively.  The presence of these structures does not allow the separation of time-averaged and turbulent quantities. At the same time, there is a wide range of plasma collisionality  in the SOL, and properly retaining collisional effects is important for its description \cite{Jorge2018}. These elements make it challenging to extend the standard gyrokinetic approach used to study core turbulence, most often based on the separation of equilibrium and fluctuating quantities and valid in the low collisionality limit, to SOL conditions. Indeed, while significant progress has been made in order to port the gyrokinetic model to the conditions of the tokamak boundary (see, e.g., Refs. \onlinecite{Qin2007, Hahm2009, Jorge2019,Frei2020}), as well as in the numerical implementation of the gyrokinetic model in the SOL geometry (see, e.g., Refs. [\onlinecite{Chang2009a,Shi2017,Pan2018}]), the numerical cost of gyrokinetic simulations of the tokamak boundary remains prohibitive and the modelling of the SOL region most often relies on fluid models  \cite{Ricci2012a,Dudson2009,Tamain2009,Easy2014,Halpern2016a,Madsen2016,Zhu2017,Paruta2018}.
The SOL fluid models, typically based on a drift-reduced set of Braginskii equations (see, e.g., Ref. \onlinecite{Zeiler1997,Scott1997}) or on a gyrofluid model (see, e.g., Ref. \onlinecite{Madsen2013}) in order to include finite Larmor radius effects, assume low plasma temperatures (and associated high plasma collisionalities) such that scale lengths are longer than the typical mean free path and deviations from a local Maxwellian distribution are small. 
However, kinetic effects might play an important role in the SOL. This is particularly true in the high confinement mode regime, when the edge temperature rises considerably and edge localised modes can become unstable, leading to the presence of high-temperature low-collisionality plasmas in the SOL \cite{Lonnroth2006,Leonard2014}. 
In the present Paper, we deduce a model for the SOL plasma dynamics that, while being able to retain the proper kinetic effects, has the potential of describing the SOL at a reduced cost with respect to full gyrokinetic simulations. 

We take advantage of the fact that, according to experimental results \cite{Endler1995,Agostini2011,Carralero2014}, SOL turbulence typically occurs on scale lengths that are larger than the ion sound Larmor radius, $\rho_s= c_s /\Omega_{ci}$ with $c_s^2 = T_e/m_i$ the sound speed and $T_e$ the electron temperature, and identify the small parameter
\begin{equation}\label{ord1}
    \epsilon \sim k_\perp \rho_s  \ll 1,
\end{equation}
where $k_\perp \sim \nabla_\perp \log \phi \sim \nabla_\perp \log n \sim \nabla_\perp \log T_e$ (while keeping $k_\perp L_p \sim 1$) with $\phi$ the electrostatic potential and $n$ the electron density.
In addition, we observe that typical turbulent time scales are ordered as
\begin{equation}\label{ord2}
    \frac{\omega}{\Omega_{i}} \sim \epsilon ^2,
\end{equation}
with $\omega \sim \partial_t \log \phi \sim \partial_t \log n \sim \partial_t \log T_e$, and the ion collision frequencies as
\begin{equation}\label{ord3}
    \frac{\nu_{i}}{\Omega_{i}} \sim \epsilon^2,
\end{equation}
ensuring that the plasma remains magnetised \cite{Frei2020}.

Based on the ordering in \cref{ord1,ord2,ord3}, a drift-kinetic (DK) model valid up to $O(\epsilon^2)$ was developed to study the plasma dynamics in the SOL in Ref. [\onlinecite{Jorge2017}].
By including the presence of large amplitude fluctuations and a full Coulomb collision operator, the model in Ref. [\onlinecite{Jorge2017}] states the evolution of the guiding-center distribution function of the plasma particles of species $a$, $F_a ({\bf R}, v_\parallel, \mu, \theta)$, where ${\bf R}$ is the particle guiding-center position, $v_\parallel = {\bf v} \cdot {\bf b}$ the velocity parallel to the magnetic field with $\bf v$ the particle velocity, ${\bf b}= {\bf B}/B$ the magnetic field unit vector, $\mu$ the magnetic moment and $\theta$ the particle's gyroangle.
A numerical efficient implementation of the DK model was then derived by expanding the $v_\parallel$ and $\mu$ dependence of the distribution function on a Hermite and Laguerre polynomial basis, respectively.
By projecting the DK equation on a Hermite-Laguerre basis, the  kinetic equation was ported to a coupled set of three-dimensional equations that describe the evolution of the moments of $F_a$. The approach was then generalised to include gyrokinetic fluctuations in Ref. \onlinecite{Frei2020}.

While the model in Ref. [\onlinecite{Jorge2017}] relies on a polynomial description of the parallel and perpendicular velocity dependencies of the distribution function, recent studies of magnetized plasma systems \cite{Mandell2018,Jorge2018,Jorge2019,Jorge2019a} point out that the $v_\parallel$ dependence may require a more accurate description than $\mu$.
Indeed the linear study\cite{Jorge2018} of the drift-wave instability using a full Coulomb collision operator shows that considerably fewer moments are necessary along the $\mu$ than the $v_\parallel$ direction to correctly estimate the linear growth rate of this instability.
The need of a refined kinetic description of the plasma in the direction parallel to the magnetic field rises also by the need to properly describe the heat conductivity in the parallel direction, since this has an important impact on the evaluation of the heat flux on the vessel walls\cite{Stangeby2000}.
In addition, the sheath dynamics might introduce a discontinuity of the distribution function particularly in the parallel direction, where the $v_\parallel$ dependence of the electron distribution function at the entrance of the magnetic pre-sheath might be discontinuos  \cite{Loizu2011,Omotani2015,Geraldini2018}. As a consequence, while a description based on a basis expansion may be particularly efficient  along the $\mu$ direction, as a low number of moments might be needed, it is worth seeking different approaches to represent the parallel dynamics. 

In the present Paper, we leverage the DK model developed in Ref. [\onlinecite{Jorge2017}] and propose an alternative approach to the solution of the DK equation.
We retain the Laguerre expansion of $F_a$ along the $\mu$ direction, while leaving $v_\parallel$ as an independent variable. 
The DK equation is then ported to a set of four-dimensional equations in the four-dimensional $(\bv{R}, v_\parallel)$ space for the perpendicular moments of $F_a$, more precisely for the coefficients of the Laguerre expansion of $F_a$.
Rather than a decomposition on a polynomial basis such as in Ref. [\onlinecite{Jorge2017}], the $v_\parallel$ dependence of the distribution function can then be treated using different numerical approaches such as finite difference, volume, or element methods.
We also express the collision operator in the kinetic equation for the guiding-center distribution function, as well as Poisson’s equation, as a function of the same set of perpendicular velocity moments.

This paper is organised as follows.
After the Introduction, \cref{solmodel} recalls the main elements of the DK model introduced in Ref. [\onlinecite{Jorge2017}].
The perpendicular moment expansion is then applied to the collisionless part of the DK equation in \cref{chap:perp_mom}. The Coulomb collision operator is introduced and expanderd in in perpendicular moments in  \cref{coul_coll_op}. \cref{chap:DK_maxwell} discusses Poisson's equation coupled to the solution of the kinetic equation.
The Conclusions follow.
In \cref{app:LB_coll_op}, the anisotropic version of the simplified Dougherty collision operator is derived.
Finally, in \cref{appendix_coeff,appendix:lag_integrals} the analytical expressions needed to evaluate the Coulomb collision operator and its moments are presented.

\section{Drift-Kinetic Model for the scrape-off layer}
\label{solmodel}

We briefly recall the main elements of the DK model derived in Ref. [\onlinecite{Jorge2017}] to study the SOL dynamics.
We first state the main assumptions behind the DK model, we then derive the DK description of single particle motion and, finally, we state the DK Boltzmann equation.

While we use the ordering in \cref{ord1,ord2,ord3}, we allow for fluctuations of $\phi$ comparable to the electron temperature by ordering
\begin{equation}
    \frac{e\phi}{T_e} \sim 1.
\label{ord4}
\end{equation}
We note that, from \cref{ord1,ord2}, the $\bv{E} \times \bv{B}$ drift, $\bv{v}_E = \bv{E}\times \bv{B}/B^2$ with ${\bf E} = -\nabla \phi$, is small with respect to  $c_s$, i.e., ${\left| {\bf v}_E \right|}/{c_s} \sim \epsilon$.
In addition, we assume that the typical turbulent time scales are comparable to the time scales associated with the $\bf E \times \bf B$ flow and to the ones of the parallel flows $v_\parallel \sim c_s$, therefore obtaining
\begin{equation}\label{ord5}
    \omega \sim k_\perp |{\bf{v_E}}| \sim k_\parallel c_s,
\end{equation}
where $k_\parallel \sim \nabla_\parallel \phi \sim \nabla_\parallel n \sim \nabla_\parallel T_e$ is the parallel wave-vector, which can be related to its perpendicular counterpart via
\begin{equation}
\frac{k_\parallel}{k_\perp} \sim \epsilon.
\label{ord6}
\end{equation}

An ordering for the electron collision frequency can be derived using \cref{ord3} and the relation $\nu_i \sim \sqrt{m_e/m_i} (T_e/T_i)^{3/2} \nu_e$, yielding
\begin{equation}\label{ord7}
    \frac{\nu_e}{\Omega_e} \sim \sqrt{\frac{m_e}{m_i}} \left( \frac{T_i}{T_e} \right)^{3/2} \epsilon^2.
\end{equation}
We remark that the ion and electron temperatures are typically comparable in the SOL, i.e., $T_i/T_e \sim 1$ \cite{Mosetto2015}. This allows us to order ${\nu_e}/{\Omega_e} \sim \sqrt{m_e/m_i} \epsilon^2$.
Finally, electromagnetic fluctuations are neglected, which restricts the present model to the case of $\beta=8\pi n T_e/B^2 \ll 1$, as well as frequencies below the shear Alfvén frequency.

We now turn to the equations of motion for a single plasma particle within the DK approximation.
We start with the Lagrangian of a single particle of species $a=\{e, i\}$ in the presence of  an electromagnetic field
\begin{align}
 L_a(\bv{x}, \bv{v})=\left[q_a \bv{A}(\bv{x}) + m_a \bv{v}\right]\cdot \dot{\bv{x}}-\left(\frac{m_a v^2}{2}+q_a \phi(\bv{x})\right).
    \label{eq:lagrangian}
\end{align}
In order to take advantage of the DK ordering, we perform a coordinate transformation from the phase-space coordinates $(\bf x, \bf v)$ to the guiding-center coordinates $({\bf R}, v_\parallel, \mu, \theta)$.
For this purpose, we introduce the right-handed set of orthonormal vectors $(\bv{e}_1, \bv{e}_2, \bf{b})$ and write the particle velocity $\mathbf v$ as
\begin{equation}
    \bv{v} = \bv{U} + \bv{v}_{\perp}'
\label{v_dec}
\end{equation}
where $\bv{U}=v_{\parallel} \bf{b} + \bv{v_E}$ and $ \bv{v}_{\perp}'=v_{\perp}' (- \sin \theta \bv{e}_1 + \cos \theta \bv{e}_2)$ with $\theta$ the particle gyroangle.
The guiding-center position $\mathbf R$ is defined as
\begin{align}
\bv{R} = \bv{x} - \bm \rho_a,
\label{rgc}
\end{align}
where ${\bm \rho}_a= |\bm \rho_{a}| (\bv{e}_1 \cos \theta + \bv{e}_2 \sin \theta)$ the particle Larmor radius with $|\bm \rho_{a}|=\sqrt{2 m_a \mu /(q_a^2 B)}$ and $\mu={m_a v_{\perp}'^{2}}/{2 B}$ the magnetic moment.

We now expand the electrostatic potential $\phi$ around $\mathbf R$ to first order in $\epsilon$, by using the guiding-center transformation in \cref{rgc}, yielding
\begin{align}
 \phi (\bv{x}) = \phi (\bv{R}) + {\bm \rho}_a \cdot \nabla_{\bv{R}} \phi (\bv{R}) + O(\epsilon^2).
\end{align}
A similar expansion procedure is applied to the magnetic vector potential ${\bf A}$.
The turbulent and gyromotion time scales are then decoupled by defining the gyroaverage operator $\left<\chi\right>$ acting on a quantity $\chi$, as
\begin{equation}\label{gyro_op}
\langle \chi \rangle = \int_0^{2 \pi} \chi (\theta) \frac{\mr{d}\theta}{2 \pi},
\end{equation}
where the integration is made at constant $\bv{R}$.
The gyroaverage operator in Eq. (\ref{gyro_op}) is applied to the Lagrangian in \cref{eq:lagrangian} yielding, up to $O(\epsilon)$,
\begin{equation}
\langle L_{a} \rangle = q_a \bv{A}^* \cdot \dot{\bv{R}}- q_a \phi^* - \frac{m_a v_{\parallel}^2}{2} + \mu \frac{m_a \dot{\theta}}{q_a} \label{lag_gc2},
\end{equation}
In \cref{lag_gc2}, we introduce the effective vector, $\bv{A}^*$, and scalar, $\phi^*$, potentials as
\begin{align}
\bv{A}^* &= \bv{A} + \frac{m_a}{q_a} \left( v_{\parallel} \hv{b} + \bv{v}_E \right),
\end{align}
and
\begin{align}
\phi^* &= \phi + \frac{m_a}{q_a} \frac{v_E^2}{2} + \frac{\mu B}{q_a} \label{lag_gc2_PR},
\end{align}
respectively.
The term $v_{E}^2 = \bv{v}_{E} \cdot \bv{v}_{E} $ in \cref{lag_gc2_PR}, although formally being $O(\epsilon^2)$, is retained since  $v^2_E \sim \epsilon^2 \Lambda^2 c_s^2$ with $\Lambda = \log \sqrt{m_i/(m_e 2 \pi)}>1$ due to the sheath boundary conditions in the SOL that set $e \phi \sim \Lambda T_e$.

The equations of motion for the guiding-center coordinates are obtained by applying the Euler-Lagrange equations to the guiding-center Lagrangian in \cref{lag_gc2}. We derive for the guiding-center velocity
\begin{align}
    \dot{\bv{R}}  &= \bv{U}+\frac{\bv{B}}{\Omega_a B_\parallel^*}\times\left(\frac{d \bv{U}}{dt}+\frac{\mu\nabla B}{m_a}\right),
    \label{eq:GC1}
\end{align}
and for the parallel acceleration
\begin{align}
    m_a \dot v_\parallel &= q_a E_\parallel - \mu \nabla_\parallel B + m_a \bv{v}_E \cdot \frac{d \hv{b}}{dt}-m_a\mathcal{A},
    \label{eq:GC2}
\end{align}
together with $\dot \theta = \Omega_a$ and $\dot \mu = 0$.
In \cref{eq:GC1,eq:GC2}, we define the convective derivative as $\mr{d}/\mr{d}t \equiv \partial_t + \bv{U} \cdot \nabla$ and the modified magnetic field $\bv{B}^*$ as $\bv{B}^* = \nabla \times \bv{A}^*$, with its parallel projection given by
\begin{equation}
    B_{\parallel}^* = \hv{b} \cdot \bv{B}^*   = B + \frac{m_a}{q_a}\hv{b} \cdot \nabla \times \left(v_\parallel \hv{b} + \bv{v}_E\right).
\end{equation}
The quantity $\mathcal{A}$ in \cref{eq:GC2} contains the higher-order nonlinear terms that ensure phase-space conservation and the Hamiltonian character of \cref{eq:GC1,eq:GC2}
\begin{equation}
    \mathcal{A}=\frac{B}{B_\parallel^*}\left(\left.\frac{d \bv{U}}{dt}\right|_\perp + \mu \nabla_\perp B\right)\cdot \frac{\nabla \times \bv{U}}{\Omega_a}.
\end{equation}

Having deduced the motion of a single particle, we now turn to their collective description. As a starting point, we note that the distribution function $f_a(\bv{x},\bv{v})$ of particle species $a$ evolves according to the Boltzmann equation, which can be written as
\begin{equation}\label{boltzmann}
\frac{\partial f_a}{\partial t} +  \bv{v} \cdot \frac{\partial f_a}{\partial \bv{x}} + \frac{q_a}{m_a}\left( \bv{E} + \frac{\bv{v} \times \bv{B}}{c} \right) \cdot \frac{\partial f_a}{\partial \bv{v}}  = C_a(f_a) ,
\end{equation}
where $C_a(f_a)= \sum_b C_{ab} (f_a, f_b)$ is the collision operator, with the summation over $b$ carried over all the particle species.
In order to write Boltzmann's equation, \cref{boltzmann}, in guiding-center coordinates, we define the guiding-center distribution function $F_a$ as
\begin{equation}\label{F_a_func}
F_a(\bv{R}, v_{\parallel}, \mu, \theta, t) = f_a[\bv{x}(\bv{R}, v_{\parallel}, \mu, \theta), \bv{v}(\bv{R}, v_{\parallel}, \mu, \theta),t],
\end{equation}
and apply the chain rule to express the derivatives in \cref{boltzmann} in terms of guiding-center variables so as to obtain
\begin{equation}
     \frac{\partial F_a}{\partial t} +\dot{\bv{R}}\cdot \nabla F_a + \dot{v_\parallel}\frac{\partial F_a}{\partial v_\parallel} + \Omega_a \frac{\partial F_a}{\partial \theta} = C_a(F_a).
     \label{eq:boltzmannSS}
\end{equation}
Finally, we apply the gyroaveraging operator to Eq. (\ref{eq:boltzmannSS}), yielding
\begin{equation}\label{drift-kinetic}
\frac{\partial \langle F_a \rangle }{\partial t} +  \dot{\bv{R}} \cdot \nabla \langle F_a \rangle + \dot{v}_{\parallel} \cdot \frac{\partial \langle F_a \rangle }{\partial v_{\parallel}}  = \langle C_a(F_a) \rangle.
\end{equation}

The right-hand side of \cref{drift-kinetic} can be further simplified by splitting the distribution function into a gyrophase dependent $\tilde F_a$ and independent $\langle F_a \rangle$ parts as $F_a=\langle F_a \rangle+\tilde F_a$, and ordering $\tilde F_a$ by subtracting \cref{drift-kinetic} from \cref{eq:boltzmannSS}. Estimating the size of each term in the resulting expression, one obtains 
 $\tilde F_a \simeq \epsilon^2 \langle F_a \rangle$ for both electrons and ions \cite{Jorge2017}. This  allows us to neglect the gyrophase dependent part of the distribution function in the collision term $C(F_a)$ and write the DK equation as
\begin{equation}\label{drift-kinetic_fin}
\frac{\partial \langle F_a \rangle }{\partial t} +  \dot{\bv{R}} \cdot \nabla \langle F_a \rangle + \dot{v}_{\parallel} \cdot \frac{\partial \langle F_a \rangle }{\partial v_{\parallel}}  = \langle  C_a(\langle F_a \rangle)\rangle.
\end{equation}
%
%
%
%

\section{Perpendicular Moment Expansion of the distribution function}\label{chap:perp_mom}

In this section, we focus on the left-hand side of the DK equation, \cref{drift-kinetic_fin}. We introduce a polynomial expansion of the distribution function $\langle F_a \rangle$ for the variable $\mu$ that allows us to port the DK equation into a set of four dimensional equations in the variables ${\bv{R}}$ and $v_\parallel$, hereby denoted as moment-hierarchy.
We obtain the recursion relation associated with this set of equations by performing an expansion of the distribution function in terms of Laguerre polynomials, $L_j (x)$, defined via the Rodrigues' formula $L_j(x) =(e^x/j!)d^j(e^{-x} x^j)/dx^j$. The Laguerre polynomials $L_j$ satisfy the recursion relation
\begin{equation}
    (j+1) {L}_{j+1} (x) = (2j +1 -x){L}_{j} (x) - j {L}_{j-1} (x),
\label{eq:laguerre1}
\end{equation}
while their derivatives can be computed using $x {{d}{L}_{j} (x)}/dx  = j [ {L}_{j} (x) - {L}_{j-1} (x)]$.
The use of Laguerre polynomials is of interest because the functions $L_j(\mu B/T)$ are orthogonal over the interval $[0,\infty)$ with respect to a Maxwellian weighting function of the form
\begin{equation}\label{orthf0}
f_a^0 = \frac{N_a e^{- s_{\perp a}^2}}{\pi  v_{th \perp a}^2},
\end{equation}
via the orthogonality relation
\begin{align}\label{orth_lag}
\int_0^{\infty} e^{-x} \pol{L}_j (x) \pol{L}_{j'} (x) \mr{d}x = \delta_{j,j'}.
\end{align}
In \cref{orthf0}, the normalized perpendicular velocity $s_{\perp a}^2$ is defined as
\begin{align}
s_{\perp a} &= \frac{v_{\perp}'}{v_{th \perp a}}=\sqrt{\frac{\mu B}{T_{\perp a}}},
\end{align}
with $v_\perp'$ the perpendicular velocity defined in \cref{v_dec} and $T_{\perp a}$ the perpendicular temperature 
\begin{align}
  T_{\perp a} &= \frac{1}{N_a} \int v_{\perp}'^2 \langle F_a \rangle \pi B dv_\parallel d\mu. \label{T_perp_def}
\end{align}
We also define the normalized parallel shifted velocity,
\begin{align}\label{shifted_vel}
s_{\parallel a} &= \frac{v_{\parallel}- u_{\parallel a}}{v_{th \parallel a}},
\end{align}
with $v_{th \parallel a}^2=2 T_{\parallel a}/m_a$, the parallel temperature
\begin{align}
  T_{\parallel a} &= \frac{1}{N_a} \int (v_{\parallel}- u_{\parallel a})^2 \langle F_a \rangle {2 \pi B}{}dv_\parallel d\mu, \label{T_par_def}
\end{align}
the parallel fluid velocity
\begin{align}
u_{\parallel a} = \frac{1}{N_a} \int v_{\parallel} \langle F_a \rangle \frac{2 \pi B}{m_a}dv_\parallel d\mu, \label{u_def}
\end{align}
and the guiding-center particle density
\begin{align}
  N_a = \int \langle F_a \rangle \frac{2 \pi B}{m_a}dv_\parallel d\mu. \label{n_a}
\end{align}

The guiding-center distribution function $\langle F_a \rangle$ is then expanded in a Laguerre basis as
\begin{equation}\label{lag_exp}
\langle F_a \rangle = f_a^0 \sum_{j=0}^{\infty} N_a^j (\bv{R}, v_{\parallel}, t) \pol{L}_j (s_{\perp}^2).
\end{equation}
where, using the ortogonality relation in \cref{orth_lag}, the coefficients $N_a^j$ can be computed via
\begin{align}
N_a^j = \frac{1}{N_a} \int \pol{L}_j (s_{\perp a}^2) \langle F_a \rangle \frac{2 \pi B}{m_a} \mr{d}\mu,
\end{align}

The coefficients $N_a^j$ can be expressed by introducing the $j$-th perpendicular moment $\lVert \chi \rVert^j_a $ of a quantity $\chi =\chi ({\bf R}, \mu, v_\parallel)$, defined as
\begin{equation}\label{mom_op}
\lVert \chi \rVert^j_a = \int \langle F_a \rangle \chi \pol{L}_j \frac{2\pi B }{m_a} \mr{d} \mu,
\end{equation}
via $N_a^j=\lVert 1 \rVert^j_a/N_a$.
%
%
%
Using this notation, the low order fluid moments $N_a, u_{\parallel a}, T_{\parallel a}$ and $T_{\perp a}$ can then be written as $N_a = \int_0^\infty \lVert 1 \rVert_a^0  \mr{d} v_{\parallel}$, $N_a u_{\parallel a}= \int_0^\infty \lVert v_{\parallel} \rVert_a^0 \mr{d} v_{\parallel}$, $N_a T_{\parallel a} = m_a \int_0^\infty \lVert (v_{\parallel}-u_{\parallel})^2 \rVert_a^0 \mr{d} v_{\parallel}$ and $N_a T_{a\perp} =\int_0^\infty \lVert \mu B \rVert_a^0 \mr{d} v_{\parallel}$, respectively.

We now derive the set of equations that state the evolution of the $N_a^j$ moments. This is a recursion relation that we denote as moment hierarchy.
As a first step, we rewrite the equations of motion, \cref{eq:GC1,eq:GC2}, in terms of the $s_{\parallel a}$ and $s_{\perp a}^2$ variables. This yields
\begin{align}
    \dot{\bv{R}} &= \bv{U}_{0 a} + \bv{U}_{p a}^* + s_{\perp a}^2 \bv{U}_{\nabla B a}^* + s_{\parallel a}^2 \bv{U}_{k a}^* + s_{\parallel a}(v_{th\parallel a} \bv{b} + \bv{U}_{p a}^{*th}),\label{eq:rdotGCform} 
\end{align}
and 
\begin{align}
    m_a \dot{v}_\parallel &= F_{\parallel a}-s_{\perp a}^2 F_{M a} +s_{\parallel a} F_{p a}^{th}-m_a \mathcal{A}. \label{eq:vparGCform}
\end{align}
In \cref{eq:rdotGCform}, the lowest-order fluid velocity $\bv{U}_{0 a} = \bv{v}_E + u_{\parallel a} \hv{b}$, and the fluid $\nabla$B drift $\bv{U}_{\nabla B a}^* = ({T_{\perp a}}/{m_a}) {\hv{b} \times \nabla B}/{\Omega_a^{*} B}$ are introduced, as well as the fluid curvature drift $\bv{U}_{ka}^{*} = ({2 T_{\parallel a}}/{m_a}) {\hv{b} \times \bv{k}}/{\Omega_a^*}$, with $\bv{k}= \hv{b} \cdot \nabla \hv{b}$, the fluid polarization drift $\bv{U}_{pa}^* = ({\hv{b}}/{\Omega_a^*}) \times {d_0 \bv{U}_{0 a}}/{dt}$, and the thermal polarization drift $\bv{U}_{p a}^{*th} = v_{th\parallel a}({\hv{b}}/{\Omega_a^*})\times (\hv{b} \cdot \nabla \bv{v}_E + \bv{v}_E \cdot \nabla \hv{b} + 2 u_{\parallel a} \bv{k})$, where $\Omega_a^{*} = q_a B_{\parallel}^* / m_a$ and ${d_{0 a}}/{dt} = \partial_t + \bv{U}_{0 a} \cdot \nabla$
In Eq (\ref{eq:vparGCform}), we introduce the parallel electric force $F_{\parallel a} = q_a E_\parallel+m_a\bv{v}_E \cdot {d_0 \hv{b}}/{dt}$, as well as the mirror force $F_{M a} = T_{\perp a} \nabla_\parallel \ln B$ and the thermal polarization force $F_{p a}^{th} =  m_a v_{th\parallel a} \hv{b} \cdot{\bv{k} \times \bv{E}}/{B}$.

The moment-hierarchy equation is obtained by projecting the DK equation, \cref{drift-kinetic_fin}, on the Laguerre polynomials $L_j$ polynomials, having expressed the distribution function according to \cref{lag_exp} and using the orthogonality relation in \cref{orth_lag}. This yields
\begin{equation}
\begin{split}
    &\frac{\partial N_a^j}{\partial t}+ \dot {\bv R}_0 \cdot \nabla N_a^j + \dot v_{\parallel 0}\frac{\partial \nabla N_a^j}{\partial v_\parallel}+F_a^j\nonumber\\
    &+ \sum_{l=j-1}^{j+1}M_{1l}^{j}\left(\dot v_{\parallel 1}\frac{\partial N_a^l}{\partial v_\parallel}+{\bv U}_{\nabla B a}^{*}\cdot \nabla N_a^l\right)=C_a^j,
\end{split}\label{momenthierarchy}
\end{equation}
with $\dot{\bv R_0} = \dot{\bv R}-s_{\perp a}^2 {\bv U}_{\nabla B a}^{*}$ the $v_\perp$ independent part of the guiding-center velocity, $\dot v_{\parallel _0}=\dot v_\parallel- s_{\perp a}^2 \dot v_{\parallel 1}$ the $v_\perp$ independent part of the parallel acceleration and $\dot v_{\parallel 1}=-F_{Ma}/m_a-(T_{\perp a}/B_{\parallel}^*)\nabla_\perp B \cdot \nabla \times {\bv U}/\Omega_a$.
Furthermore, we have introduced in \cref{momenthierarchy} the fluid term $F_a^j$ given by
\begin{align}
        F_a^j&=\sum_{l} N_a^l\frac{d_{l}^{j}}{dt}\left(\frac{N_a}{B}\right)+\left[j(N_a^j-N_a^{j-1})\left( \frac{\partial}{\partial t} +  \dot {\bv R_0} \cdot \nabla \right) \right. \nonumber \\
        &\left.  +\sum_l M_{2l}^{j}N_a^l {\bv U}_{\nabla B a}^{*}\cdot \nabla\right]\ln \left(\frac{T}{B}\right),
\label{eq:faj}
\end{align}
where the convective derivative $d_l^{j}/dt$ is defined as
${d_{l}^{j}}/{dt}=\delta_{l,j}{\partial_t}+\delta_{l,j} \dot {\bv R_0} \cdot \nabla + M_{1l}^{j}{\bv U}_{\nabla B a}^{*}\cdot \nabla$ and the perpendicular phase mixing terms as
\begin{equation}
    M_{1l}^{j}=(2j+1)\delta_{l,j}-(j+1)\delta_{l,j+1}-j \delta_{l,j-1},
\end{equation}
and
\begin{equation}
\begin{split}
    M_{2l}^{j}&=-(j+1)^2\delta_{l,j+1}+(3j^2+3j+1)\delta_{l,j}\\
    &-3j^2 \delta_{l,j-1}+j(j-1)\delta_{l,j-2}.
\end{split}
\end{equation}
Finally, the collision term $C_a^j$ is defined as
\begin{equation}
    C_a^j = \frac{1}{N_a}\int \langle C_a(\langle F_a \rangle)\rangle L_j(s_{\perp a}^2) \frac{2\pi B}{m_a} d\mu.
\label{collmoments}
\end{equation}

We note that, due to the presence of the phase-mixing terms $M_{1l}^{j}$ and $M_{2l}^{j}$, the evolution equation for the $j$-th moment $N_a^j$ is coupled its lower $N_a^{j-2},N_a^{j-1}$ and higher order $N_a^{j+1}$ counterparts.
Such coupling results from the terms containing the parallel and perpendicular gradients of the magnetic field strength $B$ in the guiding-center equations of motion, \cref{eq:GC1}, and from finite temperature gradients in \cref{eq:faj}.
%

\section{Coulomb Collision Operator}\label{coul_coll_op}

The Coulomb (or Landau) collision operator is a collision operator of the Fokker-Planck type, derived from first principles and valid in a wide range of plasma parameters, where small-angle Coulomb collisions are dominant. This operator can be written as $C_a(f_a)=\sum_b C_{ab}(f_a, f_b)$, where \cite{Rosenbluth1957}
\begin{align}
C_{ab} =\sum_{i,j=1}^3\frac{\gamma_{ab}}{2} \frac{\partial}{\partial v_i} \left[ \frac{\partial}{\partial v_j} \left( f_a \frac{\partial^2 G_b}{\partial v_i \partial v_j}  \right) - 2 \left( 1+ \frac{m_a}{m_b} \right) f_a \frac{\partial H_b}{\partial v_i}  \right], \label{RMJ_form}
\end{align}
with $\gamma_{ab} \equiv 4 \pi Z_{a}^2 Z_{b}^2 \ln \Lambda/m_a^2$ where $\ln \Lambda$ is the Coulomb logarithm, while $G_b$ and $H_b$ are the {Rosenbluth} potentials, defined as
\begin{align}
G_b (\bv{v}) = \int f_b (\bv{v}') \lvert \bv{v} - \bv{v}' \rvert \mr{d} \bv{v}',  \label{hpot}
\end{align}
and
\begin{align}
  H_b (\bv{v})= \int \frac{f_b (\bv{v}')}{\lvert \bv{v} - \bv{v}' \rvert}\mr{d} \bv{v}'. \label{gpot1}
\end{align}
The importance of retaining the full Coulomb collision operator has been shown in Refs. \onlinecite{Jorge2018,Jorge2019} by considering linear modes such as the electron plasma waves and drift waves. The growth rate and general properties of these modes might be significantly different from the ones of the Coulomb collision operator, when simplified operators are considered, in particular at typical collisionalities of the tokamak boundary.  
However,  interest in simpler operators remains, as they are able to provide the necessary diffusion in velocity space needed to perform numerical studies of low collisionality systems while satisfying basic conservation properties.
One of these operators is the anisotropic version of the Dougherty operator \citep{Hakim2020}.
This is derived in \cref{app:LB_coll_op}, together with its main conservation properties.

As a first step in the porting the Coulomb collision operator in the framework of the four-dimensional model developed herein, we note that an equivalent representation of the Coulomb collision operator can be derived from \cref{RMJ_form} by using the relationships $\nabla^2_v G_b = 2 H_b$ and $\nabla^2_v H_b = - 4 \pi f_b.$ This yields
\begin{equation}\label{collision_ji}
\begin{split}
    C_{ab} &= \frac{\gamma_{ab}}{2} \left[ \partial_{\bv{v}} \partial_{\bv{v}} f_{a} : \partial_{\bv{v}} \partial_{\bv{v}} G_{b} \right.\\
    &\left.+ 2\left(1-\frac{m_a}{m_{b}} \right) \partial_{\bv{v}} f_{a} \cdot \partial_{\bv{v}} H_{b} + 8 \pi \frac{m_a}{m_{b}} f_{a} f_{b} \right].
\end{split}
\end{equation}

%


Gyroaveraging the collision operator in \cref{collision_ji}, retaining terms up to $O(\epsilon)$ and rewriting it in terms of guiding-center coordinates, we obtain
\begin{align}
\frac{\langle C_{ab} \rangle }{\gamma_{ab}} &= \frac{2 m_a^2 \mu^2}{B^2} \frac{\partial^2 \langle F_{a} \rangle }{\partial \mu^2} \frac{\partial^2 \langle G_{b} \rangle }{\partial \mu^2} + \frac{m_a^2 \mu}{B^2} \frac{\partial^2 \langle F_{a} \rangle }{\partial \mu^2} \frac{\partial \langle G_{b} \rangle }{\partial \mu} \nonumber\\
&+ \frac{1}{2}\frac{\partial^2 \langle F_{a} \rangle }{\partial v_{\parallel}^2} \frac{\partial^2 \langle G_{b} \rangle }{\partial v_{\parallel}^2}  + \frac{m_a^2}{B^2} \frac{\partial \langle F_{a} \rangle }{\partial \mu} \frac{\partial \langle G_{b} \rangle }{\partial \mu} + \frac{4 \pi m_a}{m_{b}} \langle F_{a} \rangle \langle F_{b} \rangle \nonumber\\
&+ \frac{m_a\mu}{B} \frac{\partial^2 \langle F_{a} \rangle }{\partial v_{\parallel} \partial \mu} \frac{\partial^2 \langle G_{b} \rangle }{\partial v_{\parallel} \partial \mu} + \frac{m_a^2 \mu}{B^2} \frac{\partial \langle F_{a} \rangle }{\partial \mu} \frac{\partial^2 \langle G_{b} \rangle}{\partial \mu^2}\nonumber \\
&+ \left(1-\frac{m_a}{m_{b}} \right) \left[ \frac{2 m_a \mu}{B}\frac{\partial \langle F_{a} \rangle }{\partial \mu} \frac{\partial \langle H_{b} \rangle }{\partial \mu} +  \frac{\partial \langle F_{a} \rangle }{\partial v_{\parallel}} \frac{\partial \langle H_{b} \rangle}{\partial v_{\parallel}} \right],
\label{5Dcoll}
\end{align}
To make further progress, we  simplify the expression for $\langle C_{ab}\rangle$ by leveraging the expansion of the distribution function over an orthogonal basis. We first evaluate the Rosenbluth potentials, $G_b$ and $H_b$, and then integrate the Coulomb collision operator over $\mu$ in order to obtain an expression for the collisional moments $C_{a}^j$ in terms of moments $N_a^j$ ready to be used in the moment-hierarchy equation.

In order to perform the integrals in the Rosenbluth potentials analytically, we first rewrite $G_b$ and $H_b$ in spherical coordinates using an expansion for $f_a$ in irreducible polynomials, then performing a basis transformation to a Hermite-Laguerre polynomial basis.
Following Refs. [\onlinecite{Ji2006,Ji2008,Ji2009}], the distribution function $f_a$ is expanded in irreducible tensorial Hermite polynomials ${\bf{P}}_a^{lk}(\bf v)$ as
\begin{align}
f_a = f_{aM} \sum_{lk} \frac{{\bv{P}}_a^{lk}(\mathbf v) \cdot {\bv{M}}_a^{lk} (\bv{x},t)}{\sqrt{\sigma_k^l}} , \label{ji_held_exp} 
\end{align}
where $f_{aM}$ is the shifted Maxwellian
\begin{align}
f_{aM} = \frac{n_a e^{-s_{a}^2}}{\pi^{3/2} v_{th a}^2},
\end{align}
with $\mathbf s_a = (\mathbf v - \mathbf u_a)/v_{tha}$ the normalized shifted particle velocity, $\mathbf u_a = \int \mathbf v f_a d \mathbf v$ the fluid velocity, $v_{th a}=2 T_a/m_a$ the thermal velocity and $T_a=(T_{\parallel a}+2 T_{\perp a})/3$ the temperature.
Furthermore, we define the velocity moments  ${\bv{M}}_a^{lk}$ as
\begin{align}
{\bv{M}}_a^{lk} = \frac{1}{n_a \sqrt{\sigma_k^l}} \int \mr{d} \bv{v} {\bv{P}}_a^{lk} f_a,
\label{Irriducibile}
\end{align}
In \cref{Irriducibile}, $\sigma_k^l$ is a normalization factor
\begin{align}
\sigma_k^l = \frac{l! (l+k+1/2)!}{2^l (l+1/2)!k!},
\end{align}
and the polynomials ${\bv{P}}^{lk}$ are defined as
\begin{align}
{\bv{P}}^{lk} (\bv{v}) = \pol{L}_{k}^{l+1/2} (v^2) {\bv{P}}^l (\bv{v}), \label{tens_hermite}
\end{align}
where $\pol{L}_{k}^{l+1/2}$ are the generalized (associated) Laguerre polynomials\cite{Abramowitz1972}, given by
\begin{align}
\pol{L}_{k}^{l+1/2} (x) = \sum_{m=0}^{k} L^l_{km} x^m, 
\end{align}
with coefficients
\begin{align}
L^l_{km} = \frac{(-1)^m (l+k+1/2)!}{(k-m)! (l+m+1/2)! m!}. \label{gen_lag_def}
\end{align} 
and ${\bv{P}}^{l} (\bv{v})$ are the totally symmetric and traceless tensors, defined as
\begin{equation}\label{sym_tens}
{\bv{P}}^l(\bv{v}) = \frac{(-1)^l v^{2l+1}}{(2l-1)!!}\left( \frac{\partial}{\partial \bv{v}}\right)^l \frac{1}{v}.
\end{equation}

In order to analytically compute the integrals present in the Rosembluth potentials, $G_b$ and $H_b$, we expand the function $\lvert \bv{v}- \bv{v}' \rvert^{-1}$ in terms of Legendre polynomials $P_l(x)=[d^l(x^2-1)^l/dx^l]/(2^l l!)$ as
\begin{align}
\frac{1}{\lvert \bv{v}- \bv{v}' \rvert} = \frac{1}{\sqrt{v^2 + v'^2 - 2 v v' \xi'}} = \sum_{l=0}^{\infty} \frac{v_{<}^l}{v_{>}^{l+1}} \pol{P}_l (\xi'), \label{Leg_exp}
\end{align}
where $v_{<} = \mr{min} (v, v') $ and $v_{>} = \mr{max} (v, v')$, while $\xi' = \bv{v} \cdot \bv{v}'/ ( \lvert \bv{v} \rvert \vert \bv{v}' \rvert ) $ is the cosine of the angle between the vectors $\bv{v}$ and $\bv{v}'$.
This yields for $H_b$
\begin{align}
  H_b (\bv{v})= \sum_{l',k,l}\frac{{\bf M}_a^{l'k}}{\sqrt{\sigma_k^{l'}}}\cdot\int f_{bM} {\bv P}_b^{l'k} \frac{v_{<}^l}{v_{>}^{l+1}} \pol{P}_l (\xi') v^2 d \xi' dv d\theta, \label{gpot}
\end{align}
and a similar expression for $G_b$ is obtained.
The integration over the  angle $\theta$ in \cref{gpot} is performed using the following identity\cite{Ji2006} for the irreducible polynomials ${\bv P}^l$
\begin{align}
  \int_0^{2\pi}  {\bv{P}}^l (\bv{v}')  \mr{d} \theta'_{\hv{v}} = 2 \pi v'^l \pol{P}_l (\xi') {\bv{P}}^l (\hv{v}) , \label{gyro_formula}
\end{align}
the $\xi'$ integration is performed using the orthogonality relations for the Legendre polynomials
\begin{align}
\int_{-1}^{1} \pol{P}_l (\xi') \pol{P}_n (\xi') \mr{d} \xi' = \frac{\delta_{ln}}{l+1/2},
\end{align}
and the integration over the speed variable $v$ is performed by splitting the cases $v'<v$ and $v'>v$, and defining  $I_+^k=2 \int_{0}^{s_b} \mr{d} v' v'^{k} e^{-v'^2}/\sqrt{\pi}$ and  $I_-^k=2\int_{s_b}^{\infty} \mr{d} v' v'^k e^{-v'^2}/\sqrt{\pi}$. This yields the following form for the Rosenbluth potentials
\begin{align}
  H_{b} &=  \frac{n_b}{v_{thb}} \sum_{l,k} \sum_{m=0}^{k}  \frac{L_{km}^l}{\sqrt{\sigma_k^l}} \frac{{\bv{M}}_a^{lk}
  \cdot {\bv{P}}^l(\hv{s})}{l + 1/2}  s_b^l\left( \frac{\mathrm{I}_{+}^{2(l+m+1)}}{s_{b}^{2l+1}} + \mathrm{I}_{-}^{2m+1} \right),
  \label{Hrosenbluth} \\
  G_{b} &=  n_b v_{thb} \sum_{l,k} \sum_{m=0}^{k} \frac{L_{km}^l}{\sqrt{\sigma_k^l}}  \frac{{\bv{M}}_a^{lk}
 \cdot {\bv{P}}^l(\hv{s})}{l + 1/2}  s_b^l \nonumber\\
 &\times\left[ \frac{1}{2l+3} \left( \frac{\mathrm{I}_{+}^{2(l+m+2)}}{s_b^{2l+1}} + s_b^2\mathrm{I}_{-}^{2m+1} \right) \right.\nonumber \\
 & \left. - \frac{1}{2l-1} \left( \frac{\mathrm{I}_{+}^{2(l+m+1)}}{s_b^{2l-1}} + \mathrm{I}_{-}^{2m+3} \right) \right].
 \label{Grosenbluth}
\end{align}

We now write the integrals in \cref{Hrosenbluth,Grosenbluth} in a form suitable to express the gyroaveraged Rosenbluth potentials appearing in \cref{5Dcoll} in terms of the moments $N_a^j$. For this purpose, we expand the  integrals in $I_+^{2k}$ and $I_-^{2k+1}$ in powers of $s$.
First, we Taylor-expand the integrand in $I_+^{2k}$ around $s'=s$ as
\begin{equation}\label{exp_ros}
e^{-s'^2} = e^{-s^2} \sum_{q=0}^{\infty} \frac{(s^2- s'^2)^q}{q!},
\end{equation}
yielding
\begin{align}
I_{+}^{2k} = \frac{2e^{-s^2}}{\sqrt{\pi}}  \sum_{q=0}^{\infty} s^{1+2k+2q} \frac{(k-1/2)!}{2(k+q+1/2)!}. \label{i+2}
\end{align}
A similar procedure is applied to the integrand in $I_-^{2k+1}$, which is Taylor expanded around $s' = 0$, yielding
\begin{align}
  I_-^{2k+1} &= \sum_{j=0}^{k}\frac{k!}{j!}s^{2j}\frac{e^{-s^2}}{\sqrt{\pi}}. \label{i-2}
\end{align}
This method yields the following expression for the gyroaveraged Rosenbluth potentials
\begin{align}
  \langle  H_{b}(\bv{s}) \rangle =  \frac{N_b v_{thb \parallel}}{v_{thb}} \sum_{l,k =
  0}^{\infty} \mathcal{N}_b^{lk} h_{00}^{lk}, 
 \label{Rosenbluth1}
\end{align}
\begin{equation}
    \langle  G_{b}(\bv{s}) \rangle  =  N_b v_{thb} v_{thb \parallel} \sum_{l,k =
 0}^{\infty} \mathcal{N}_b^{lk} g_{00}^{lk},
 \label{Rosenbluth2}
\end{equation}

In \cref{Rosenbluth1,Rosenbluth2}, we introduce the Hermite polynomials $H_p(x)=(-1)^p \exp(x^2)d^p\exp(-x^2)/dx^p$, the fluid moments 
\begin{align}
\mathcal{N}_a^{lk}&= \frac{2^l (l!)^2 }{(2l)!(l+1/2)\sigma_k^l}  \sum_{p=0}^{l+2k} \sum_{j=0}^{k+\floor{l/2}} T^{pj}_{lk} \int_{-\infty}^{\infty}  \pol{H}_p(s_\parallel) N_a^{j} \mr{d} \mr{s}_{\parallel} . \label{mom_trans0}
\end{align}
the velocity-dependent terms 
\begin{align}
h_{00}^{lk} = \sum_{n=0}^{\infty} h^{lkn} \beta_{\perp}^{-n} s_{\perp}^{2n} e^{-\beta_{\perp}^{-1} s_{\perp}^2}, \label{hlkn}
\end{align}
and
\begin{align}
g_{00}^{lk} = \sum_{n=0}^{\infty} g^{lkn} \beta_{\perp}^{-n} s_{\perp}^{2n} e^{-\beta_{\perp}^{-1} s_{\perp}^2},\label{glkn}
\end{align}
with $\beta_\parallel=v_{thb}^2/v_{thb\parallel}^2=T_b/T_{\parallel b}$ and $\beta_{\perp}=v_{thb}^2/v_{thb\perp}^2=T_b/T_{\perp b}$
as well as the coefficients $T_{lk}^{pj}$, which allow us to convert between Hermite-Laguerre and Legendre-Laguerre polynomials via

\begin{align}
\pol{P}^l(\xi) s^l \pol{L}_k^{l+1/2}(s^2) = \sum_{p=0}^{l+2k} \sum_{j=0}^{k + \floor{l/2}}  T^{pj}_{lk} \pol{H}_p ({s_{\parallel}}) \pol{L}_j (s_{\perp}^2), \label{bas_trans2}
\end{align}
with the inverse transform given by
\begin{align}
\pol{H}_p ({s_{\parallel}}) \pol{L}_j (s_{\perp}^2) = \sum_{l=0}^{p+2j} \sum_{k=0}^{j + \floor{p/2}}  \left(T^{-1}\right)^{lk}_{pj} \pol{P}^l(\xi) s^l \pol{L}_k^{l+1/2}(s^2). \label{bas_trans1}
\end{align}
An analytically closed formula for $T^{pj}_{lk}$ and $\left(T^{-1}\right)^{lk}_{pj}$ is given in Ref. [\onlinecite{Jorge2017}].

We now derive the expression for the perpendicular moments $C_a^j$ of the Coulomb collision operator in \cref{collmoments} in terms of moments $N_a^j$ of the guiding-center distribution function.
We first rewrite the velocity derivatives of the Rosenbluth potentials $h_{00}^{lk}$ and $g_{00}^{lk}$ as
\begin{align}
\frac{\partial^{i+j} h_{00}^{lk}}{\partial s_{\parallel}^i \partial (s_{\perp}^2)^j} = \sum_{n=0}^{\infty} h^{lkn}_{ij} \beta_{\perp}^{-n} s_{\perp}^{2n} e^{-\beta_{\perp}^{-1}s_{\perp}^2},
\label{eq:hijlkn}
\end{align}
and
\begin{equation}
    \frac{\partial^{i+j} g_{00}^{lk}}{\partial s_{\parallel}^i \partial (s_{\perp}^2)^j} = \sum_{n=0}^{\infty} g^{lkn}_{ij} \beta_{\perp}^{-n} s_{\perp}^{2n} e^{-\beta_{\perp}^{-1}s_{\perp}^2}.
\label{eq:gijlkn}
\end{equation}
with the coefficients $h^{lkn}_{ij}$ and $g^{lkn}_{ij}$ given in \cref{appendix_coeff}.
The projection of the Coulomb collision operator on the Laguerre basis can then be written in the following form
\begin{align}\label{coll_op_mom}
 C_{ab}^{j}  &= \hat{\nu}_{ab} \frac{N_b}{n_b} \sum_{i=0}^2 \sum_{lkp}^{\infty} \frac{\partial^i  N^p_a(s_{\parallel a}) }{\partial s_{\parallel a}^i} \mathcal{N}^{lk}_b \mathcal{C}_i^{lkpj} \nonumber \\
& + 4 \hat{\nu}_{ab} \frac{m_a}{m_b} \sum_{np}^{\infty} N^p_a(s_{\parallel a}) N^n_b( s_{b\parallel})D_{pj}^{n}  ,
\end{align}
with $\hat{\nu}_{ab} =   \nu_{ab}v_{th b} = {\gamma_{ab} n_b v_{thb}}/{v_{tha}^3}$ and 
\begin{align}
\mathcal{C}_i^{lkpj} &= \sum_{n=0}^{\infty} \sum_{rs=0}^2  a_{i,rs}^{npj} h^{lkn}_{rs} + b_{i,rs}^{npj} g^{lkn}_{rs}, \label{a_b_coeff}
\end{align}
where the numerical coefficients $a_{i,rs}^{npj}, b_{i,rs}^{npj}$ are given by
\begin{align}
a_{0,01}^{npj} &= -\frac{4 \theta^{3/2} \beta_{\perp}}{\beta_{\parallel}^{1/2}} \left( 1 - \frac{m_a}{m_b} \right)(1+p) \left( C_{p+1,j}^n - C_{pj}^{n} \right), \\
b_{0,01}^{npj} &= - 4 \frac{\theta^{1/2} \alpha_{\perp} \beta_{\perp}}{\beta_{\parallel}^{1/2} } (p+1)C_{p+1,j}^n , \\
b_{0,02}^{npj} &= 4 \frac{\beta_{\perp}^2 \theta^{3/2}}{\beta_{\parallel}^{1/2}}  (p+1) \left\{ 2(p+2)\left( C_{p+2,j}^n - 2C_{p+1,j}^n + C_{pj}^n \right)\right.\nonumber\\
&\left.+ C_{p+1,j}^n - C_{p,j}^n  \right\}, \\
a_{1,01}^{npj} &=  \alpha_{\parallel}^{1/2} \left( 1 - \frac{m_a}{m_b} \right)C_{p+1,j}^n , \\
b_{1,11}^{npj} &= 2  \beta_{\perp} \alpha_{\parallel}^{1/2} (p+1) \left( C_{p+1,j}^n - C_{pj}^n \right), \\
b_{2,20}^{npj} &= \frac{1}{2} \theta^{1/2} \alpha_{\parallel} \beta_{\parallel}^{1/2} C_{p,j}^n.
\end{align}
In addition, the integral terms $C_{pj}^m$ and $D_{pj}^m$ that result, respectively, from the product between $F_a$ and the Rosenbluth potentials and from the product $\langle F_a \rangle \langle F_b \rangle$, are defined as
\begin{align}
C_{pj}^m  &= \int_0^{\infty} \beta_{\perp}^{-m} s_{b\perp}^{2m} \pol{L}_p (s_{\perp a}^2)  \pol{L}_j (s_{\perp a}^2)  e^{- \beta_{\perp}^{-1}s_{b\perp}^2 - s_{\perp a}^2} \mr{d} s_{\perp a}^2, \label{lag-int1}
\end{align}
and
\begin{align}
D_{pj}^m  &= \int_0^{\infty} \pol{L}_p (s_{\perp a}^2) \pol{L}_p (s_{\perp a}^2) \pol{L}_m (s_{b\perp}^2) e^{- s_{b\perp}^2 - s_{\perp a}^2} \mr{d} s_{\perp a}^2. \label{lag-int2}
\end{align}
The expressions for $C_{pj}^m$ and $D_{pj}^m$ are reported in \cref{appendix:lag_integrals}.
For convenience, the dimensionless quantities $\theta, \alpha_{\parallel,\perp}$ are introduced, which are defined as $\theta = v_{tha}^2/v_{thb}^2$, $\alpha_{\perp} = T_a/T_{a\perp }$ and as $\alpha_{\parallel} = T_a/T_{a\parallel }$.

\section{Drift-Kinetic Poisson's Equation}\label{chap:DK_maxwell}

The electric field appearing in the DK equation, \cref{drift-kinetic_fin}, and subsequently in the moment-hierarchy equation, \cref{momenthierarchy}, is evaluated using Poisson's equation, which can be written as
\begin{align}
\nabla^2 \phi &= - 4 \pi \sum_{a}  q_{a} \int f_{a} \mr{d} \bv{v}. \label{Phipotentials}
\end{align}
In order to rewrite Poisson's equation in terms of moments $N_a^j$ of the guiding-center distribution function $F_a$, we express the velocity space volume element in \cref{Phipotentials} as $d \bv{v}=\delta(\bv{x} - \bv{R} - \bv{\rho})B_\parallel^* dv_\parallel d\mu d\theta d \bv{R}/m_a$, and we integrate \cref{Phipotentials} over $\bv{R}$ and $\theta$. This allows us to rewrite the Poisson equation as
\begin{align}
\nabla^2 \phi &= - 4 \pi \sum_{a}  q_{a} \int \langle F_{a}(\bv{x}-\bv{\rho},\mu, v_\parallel, \theta)\rangle \frac{2 \pi B_{\parallel}^*}{m_a} \mr{d} v_\parallel d\mu,
\end{align}

Introducing the Fourier-transform of the distribution function $F_{ak}=F_{ak}(\bv{k},v_\parallel, \mu, \theta)$, defined via $F_a=\int d\bv{k}F_{ak}e^{-i \bv{k}\cdot\bv{R}}$, and the Jacobi-Anger expansion
\begin{equation}
e^{i \bv{k} \cdot\bv{\rho} } = J_0 (k_{\perp}\rho) + 2 \sum_{l=1}^{\infty}i^l J_l (\rho k_{\perp}) \cos  l\theta,
\end{equation}
with $i$ the imaginary unit, we obtain the following form for the Poisson's equation 
\begin{equation}
\begin{split}
    \nabla^2 \phi (\bv{x}) &= - 4 \pi \sum_{a} q_{a} \int \mr{d}v_{\parallel} \mr{d}\mu \mr{d}\theta\frac{B_{\parallel}^*}{m}\\
    &\times \left(  \Gamma_0 [ F_{ak} ]+ 2 \sum_{l=1}^{\infty}i^l \Gamma_l [ F_{ak}  \cos l\theta] \right)  .
\end{split}\label{poiss_eq_2}
\end{equation}
with the Fourier-Bessel operator $\Gamma_l \left[ f\right]$ defined as
\begin{equation}
\Gamma_l \left[ f(\bv{k}) \right] = \int  J_l ( k_{\perp} \rho) f(\bv{k}) e^{-i \bv{k} \cdot \bv{x}} \mr{d} \bv{k}.
\end{equation}
We now consider the DK limit of Poisson's equation, \cref{poiss_eq_2}.
As pointed out in Ref. \onlinecite{Jorge2017}, due to the asymptotic form of the Bessel function $J_l$ for small arguments $J_l(x) \sim x^l$, and the fact that $F_a \simeq \langle F_a \rangle + O(\epsilon^2)$, only the zeroth order function $J_0$ is needed.
Furthermore, $J_0$ can be written in terms of Laguerre polynomials by making use of the identity \cite{Mandell2018,Jorge2019,Frei2020}
\begin{align}
J_0 (k_{\perp} \rho) = \sum_{n=0}^{\infty} K_n ( k_{\perp}\rho_{th \perp a}) \pol{L}_n (s_{\perp}^2) \label{j0_dec},
\end{align} 
with $\rho_{th \perp a}=v_{th\perp a}/\Omega_a$ and $K_n$ given by
\begin{align}
K_n (\rho_{th \perp a} k_{\perp}) = \frac{1}{n!} \left( \frac{  k_{\perp}\rho_{th \perp a}}{2} \right)^{2n} e^{-\left( \frac{  k_{\perp} \rho_{th \perp a} }{2}\right)^2 }.
\label{eq:knrhok}
\end{align}
Equations (\ref{j0_dec}-\ref{eq:knrhok}) allows us to decouple the spatial dependence in $J_0$ from its velocity dependence.
Finally, noting that  $K_n(x) \sim x^{2n}$ for $x \ll 1$, we retain the $n=0$ and $n=1$ terms in \cref{j0_dec} and expand both $K_0$ and $K_1$ up to $O(\epsilon^2)$, yielding
%
%
\begin{equation}
\begin{split}
    \nabla^2 \phi (\bv{x}) &= - 4 \pi \sum_{a} q_{a} N_a \int \mr{d}v_{\parallel} \frac{B_{\parallel}^*}{B}\\
    &\times\left(N_a^0-\frac{\rho_{th \perp a}^2}{4}\nabla_\perp^2 N_a^0+\frac{\rho_{th \perp a}^2}{4}\nabla_\perp^2 N_a^1\right).
\end{split}\label{poiss_eq_4}
\end{equation}
The final form of the DK Poisson's equation is obtained by noting that $B_\parallel^*/B=1+O(\epsilon)$. This allows us to write \cref{poiss_eq_4} as
\begin{equation}
\begin{split}
    \nabla^2 \phi (\bv{x}) &= - 4 \pi \sum_{a} q_{a} N_a \int \mr{d}v_{\parallel} \\
    &\times\left(\frac{B_{\parallel}^*}{B}N_a^0-\frac{\rho_{th \perp a}^2}{4}\nabla_\perp^2 N_a^0+\frac{\rho_{th \perp a}^2}{4}\nabla_\perp^2 N_a^1\right).
\end{split}\label{poiss_eq_fin}
\end{equation}
We remark that the Poisson equation in \cref{poiss_eq_fin} reduces to the one in Ref. [\onlinecite{Jorge2017}] when the integration over $v_\parallel$ is carried out.

\section{Conclusions}

In the present work, a four-dimensional moment model suitable to describe the plasma dynamics in the SOL region of magnetic confinement fusion devices at arbitrary collisionality is derived.
The model is based on the moment-hierarchy  equation, \cref{momenthierarchy}. This equation is used to evolve the  moments of the gyroaveraged distribution function $\langle F_a \rangle$, and it is obtained by projecting the collisional DK equation, \cref{drift-kinetic_fin}, over a Laguerre basis in the perpendicular velocity space,  while $v_\parallel$ remains an independent variable of the resulting system of equations.
A description using a Laguerre polynomial basis allows us to express analytically the nonlinear Coulomb collision operator, as well as the DK Poisson's equation, in terms of perpendicular velocity moments of $\langle F_a \rangle$.
While \cref{momenthierarchy}, is written for an infinite number of moments and is valid for distribution functions arbitrarily far from equilibrium, in practice, a closure scheme must be provided in order to reduce the model to a finite number of equations.
The semi-collisional closure (see, e.g., Refs [\onlinecite{Zocco2011,Loureiro2015,Jorge2017}]) can provide the formalism to evaluate such a closure, allowing the description of the necessary kinetic effects at an arbitrary level of collisionality. 
We remark that, leveraging the work in Ref.  \onlinecite{Frei2020}, the model derived here can be used as a starting point for the development of a four-dimensional gyrokinetic moment-hierarchy.

\section{Acknowledgments}

This work has been carried out within the framework of the EUROfusion Consortium and has received funding from the Euratom research and training programme 2014-2018 and 2019-2020 under grant agreement No 633053, from Portuguese FCT (Fundação para a Ciência e Tecnologia) under grant PD/BD/105979/2014, carried out as part of the training in the framework of the Advanced Program in Plasma Science and Engineering (APPLAuSE,) sponsored by FCT under grant No. PD/00505/2012 at Instituto Superior Técnico, from the Swiss National Science Foundation and by a grant from the Simons Foundation (560651, ML).
The views and opinions expressed herein do not necessarily reflect those of the European Commission.

\section{Data Availability}

Data sharing is not applicable to this article as no new data were created or analyzed in this study.

\appendix
\addtocontents{toc}{\protect\setcounter{tocdepth}{0}}

\section{Anisotropic Dougherty Collision Operator}
\label{app:LB_coll_op}

In addition to the Coulomb collision operator, we consider here the Dougherty collision operator, a simplified collision operator that is of interest for implementation in the weakly collisional case. We generalise this operator to retain temperature anisotropy effects and we port it in the framework of the four-dimensional model developed herein.
The Dougherty operator \cite{Dougherty1964}, $C_D$, is defined as
\begin{align}
C_{D} \left( f_a \right) = \nu_a \frac{\partial }{\partial \bv{v}} \left[ \left( \bv{v} - \bv{u}_a\right) f_a +  \frac{T_a}{m_a} \frac{\partial f_a}{\partial \bv{v}} \right], \label{CLBD}
\end{align}
where $\mathbf u_a=\int \mathbf v f_a d\mathbf v/n_a$ is the fluid velocity.
It can be shown that the operator in \cref{CLBD} conserves particles, momentum and energy, satisfies an H-theorem and and vanishes if $f_a$ is a Maxwellian.
Furthermore, when written in terms of guiding-center variables $(\mathbf R, v_\parallel, \mu, \theta)$ and applied to an isotropic Hermite-Laguerre basis $H^{pj}=H_p[(v_\parallel-u_{\parallel a})/v_{tha}] L_j(\mu B/T_a)$ with $v_{tha}^2=2 T_a/m_a$, the Dougherty operator in \cref{CLBD} yields
\begin{equation}
    C_D(H^{pj})=-\nu(p+2j)H^{pj},
\end{equation}
showing that a Hermite-Laguerre polynomial basis is an eigenfunction of the Dougherty operator.

To generalise the Dougherty collision operator $C_D$ to an anisotropic Hermite-Laguerre basis $H_p(s_{\parallel a}^2) L_j(s_{\perp a}^2)$, we first rewrite  \cref{CLBD} in a covariant form, by replacing the differential operators by their covariant counterparts, yielding
\begin{align}
C_{D}  =\nu(  3f_a  + \omega^i  f_{a;i} +  D^{ij} f_{a ; i; j} \label{LB_gen}),
\end{align}
with $\bv{v} - \bv{u}_a=\bv{\omega}$ the friction vector and $D^{ij}=\delta^{ij}T_a/m_a$ the second-order covariant diffusion tensor.
The first and second covariant derivatives in \cref{LB_gen} of the scalar function $f$ are defined as
\begin{align}\label{cov_der1}
f_{;i} = \frac{\partial f}{\partial \xi_i},
\end{align}
and
\begin{equation}\label{cov_der}
    f_{; i;j} = \frac{\partial^2 f}{\partial \xi^i \partial \xi^j} - \Gamma_{ij}^k \frac{\partial f}{\partial \xi^k},
\end{equation}
respectively, with $\Gamma_{ij}^k$ the Christoffel symbols of the second kind for the new coordinate system $\bf \xi(\bf x, \bf v)$.
For the case of velocity guiding-center coordinates $(\mu, v_\parallel, \theta)$, the symbols $\Gamma_{ij}^k$ can be derived from the guiding-center metric-tensor $g_{ij}$
\begin{equation}\label{metric}
  g_{ij}=
\left[
\begin{array}{ccc}
\frac{B}{2m_a \mu} & 0 & 0 \\
  0 & \frac{2\mu B}{m_a} & 0 \\
  0 & 0 & 1 \\
\end{array}
\right],
\end{equation}
using the following expression for $\Gamma_{ij}^k$
\begin{equation}
\Gamma_{ij}^k = \frac{1}{2}g^{kl} \left( \frac{\partial g_{il}}{\partial \xi^j} + \frac{\partial g_{jl}}{\partial \xi^i} - \frac{\partial g_{ij}}{\partial \xi^l}   \right).\label{christ_symb}
\end{equation}
This yields
\begin{align}\label{gammamu}
  \Gamma^{\mu}_{ij} =
\left[
\begin{array}{ccc}
- \frac{1}{2 \mu} & 0 & 0 \\
  0 & - 2 \mu & 0 \\
  0 & 0 & 0 \\
\end{array}
\right],
\end{align}
and
\begin{align}\label{gammaphi}
\Gamma^{\theta}_{ij} =
\left[
\begin{array}{ccc}
 0 & \frac{1}{2 \mu} & 0 \\
  \frac{1}{2 \mu} & 0 & 0 \\
  0 & 0 & 0 \\
\end{array}
\right],
\end{align}
as well as $\Gamma_{ij}^{v_{\parallel}} = 0$.

To generalise the diffusion tensor $D^{ij}$ to the anisotropic case, we start by considering the following form for $D^{ij}$ in the Cartesian coordinates $(x,y,z)$
\begin{equation}
  D^{ij} = \frac{1}{m_a}
\left[
\begin{array}{ccc}
T_x & 0 & 0 \\
  0 & T_y & 0 \\
  0 & 0 & T_z \\
\end{array}
\right],
\end{equation}
where $T_{x}=\int m v_{x}^2/2 f d\bf v$, and analogous definitions apply to $T_{y}$ and $T_z$.
By identifying the $z$ axis with the direction of the magnetic field, we consider $T_x = T_y = T_{\perp}$ and $T_z = T_{\parallel}$.
By performing the coordinate transformation from Cartesian to the DK coordinates $(\mu, v_{\parallel},\theta)$, we obtain
\begin{equation}
  D^{ij} = 
\left[
\begin{array}{ccc}
\frac{2 T_{\perp} \mu }{B} & 0 & 0 \\
  0 & \frac{T_{\parallel}}{m_a} & 0 \\
  0 & 0 & \frac{T_{\perp}}{2\mu B} \\
\end{array}
\right].
\end{equation}
The anisotropic Dougherty collision operator in guiding-center coordinates can then be written as
\begin{align}
\langle C_{D} \left[ F_a \right] \rangle &= \nu \left[  3 \langle F_a  \rangle+ \left( v_{\parallel}- u_{\parallel a} \right)\frac{\partial \langle F_a  \rangle }{\partial v_{\parallel}} + 2 \mu  \frac{\partial \langle F_a  \rangle }{\partial \mu} \right. \nonumber \\
& \left. +  \frac{T_{\parallel a}}{m_a} \frac{\partial^2 \langle F_a  \rangle }{\partial v_{\parallel}^2} +  \frac{2 T_{\perp a}}{B} \frac{\partial }{\partial \mu} \left(\mu \frac{\partial \langle F_a  \rangle}{\partial \mu} \right) \right] .  \label{coll_op_LBD}
\end{align}
The collision operator defined in \cref{coll_op_LBD} conserves particle, momentum and energy. It vanishes for a bi-Maxwellian $f_{aM}$ of the form
\begin{align}
f_{aM} = \frac{n_a m_a }{\pi^{3/2} v_{th \parallel a} 2 T_{\perp a} } \exp \left( -\frac{(v_{\parallel} - u_{\parallel a} )^2}{v_{th \parallel a}^2} - \frac{v_\perp^2}{v_{th\perp a}^2} \right) \label{an_bi_max}
\end{align}
and it can be shown that it satisfies the H-theorem for a near-Maxwellian distribution.

Finally, the perpendicular moments $C_a^j$ of the anisotropic Dougherty collision operator can be derived by plugging \cref{coll_op_LBD} in \cref{collmoments}, yielding
\begin{align}
C_a^j = \nu \left[ (1-2j) N^j_a  + s_{\parallel} \frac{\partial  N^j_a  }{\partial s_{\parallel}} + \frac{1}{2} \frac{\partial^2  N^j_a   }{\partial s_{\parallel}^2} \right]. \label{LBD_moment}
\end{align}

\section{Coefficients of the Rosenbluth Potentials}
\label{appendix_coeff}
We  write the coefficients $h_{ij}^{lkn}$ and $g_{ij}^{lkn}$ of the expansion of the Rosenbluth potentials, $H$ and $G$, needed to compute the expressions in \cref{eq:hijlkn,eq:hijlkn}.
For the $H$ potential, we write 
\begin{equation}
h^{lkn}_{ij} = \sum_{u=1}^2 h^{lkn}_{iju}
\end{equation}
where
\begin{align*}
\left. h^{lkn}_{ij1} \right. &= \sum_{m=0}^k  \sum_{j=0}^{\floor{l/2}} \sum_{q=0}^{\infty} \frac{P_{jl} L_{km}^l}{\sqrt{\pi}} \frac{(m+l+1/2)!}{(m+l+q+3/2)!} \times \\
&\binom{j+q+m+1}{n} e^{-\beta_{\parallel}^{-1} s_{\parallel}^2} \times \\
&\times f_{ij}(j+q+m+1-n,l+2q+2m-2n+2) , \\
\left. h^{lkn}_{ij2} \right. &= \sum_{m=0}^k  \sum_{j=0}^{\floor{l/2}} \sum_{q=0}^{m} \frac{P_{jl} L_{km}^l}{\sqrt{\pi}} \frac{m!}{q!} \times \\
&\times \binom{j+q}{n} e^{-\beta_{\parallel}^{-1} s_{\parallel}^2} \times \\  &\times f_{ij} (j+q-n,l+2q-2n),
\end{align*}

Similarly, for the $G$ potential, we expand 
\begin{equation}
g^{lkn}_{ij}= \sum_{u=1}^4 g^{lkn}_{iju}
\end{equation}
where the coefficients $g_{iju}^{lkn}$ are given by
\begin{align*}
	\left. g^{lkn}_{ij1} \right. &= \sum_{m=0}^k \sum_{j=0}^{\floor{l/2}} \sum_{q=0}^{\infty} \frac{P_{jl} L_{km}^l}{(2l+3)\sqrt{\pi}} \frac{(m+l+3/2)!}{(m+l+q+5/2)!} \times \\
	&\times \binom{j+q+m+2}{n}e^{-\beta_{\parallel}^{-1} s_{\parallel}^2}  \times \\
	& \times f_{ij} (j+q+m+2-n,l+2q+2m-2n+4), \\
	\left. g^{lkn}_{ij2} \right. &= \sum_{m=0}^k \sum_{j=0}^{\floor{l/2}} \sum_{q=0}^{m} \frac{P_{jl} L_{km}^l}{(2l+3)\sqrt{\pi}} \frac{m!}{q!}  \times \\
	&\times \binom{j+q+1}{n} e^{- \beta_{\parallel}^{-1} s_{\parallel}^2} \times \\
	& \times f_{ij} (j+q+1-n,l+2q+2m-2n+2), \\
	\left. g^{lkn}_{ij3} \right. &= \sum_{m=0}^k  \sum_{j=0}^{\floor{l/2}} \sum_{q=0}^{\infty} \frac{P_{jl} L_{km}^l}{(1-2l) \sqrt{\pi}} \frac{(m+l+1/2)!}{(m+l+q+3/2)!}  \times \\
	&\times \binom{j+q+m+2}{n} e^{-\beta_{\parallel}^{-1} s_{\parallel}^2} \times \\
	&\times f_{ij}(j+q+m+2-n,l+2q+2m-2n+4), \\
	\left. g^{lkn}_{ij4} \right. &= \sum_{m=0}^k  \sum_{j=0}^{\floor{l/2}} \sum_{q=0}^{m+1} \frac{P_{jl}L_{km}^l(m+1)!}{(1-2l)\sqrt{\pi}q!}  \times \\
	&\times \binom{q+j}{n} e^{- \beta_{\parallel}^{-1} s_{\parallel}^2} \times \\
	& \times f_{ij} (j+q-n,l+2q+2m-2n).
\end{align*}
In the previous expressions, the function $f_{ij}$ is introduced
\begin{align*}
f_{00} (x,y) &= \frac{s_{\parallel}^{y}}{\beta_{\parallel}^{y/2}}, \\
f_{01} (x,y) &= \frac{s_{\parallel}^{y-2}}{\beta_{\parallel}^{y/2-1}}\left[x - \frac{s_{\parallel}^2}{\beta_{\parallel}}\right], \\ 
f_{10} (x,y) &= \frac{s_{\parallel}^{y-1}}{\beta_{\parallel}^{y/2} } \left[y - 2\frac{s_{\parallel}^2}{\beta_{\parallel}} \right] ,\\
f_{11} (x,y) &= \frac{s_{\parallel}^{y-3}}{\beta_{\parallel}^{y/2-1} \beta_{\perp}} \left[x(y-2) - (2x+y)\frac{s_{\parallel}^2}{\beta_{\parallel}} + 2 \frac{s_{\parallel}^4}{\beta_{\parallel}^2}\right]  , \\
f_{02} (x,y) &= \frac{s_{\parallel}^{y-4}}{\beta_{\parallel}^{y/2-2} \beta_{\perp}^2} \left[ \left(x - \frac{s_{\parallel}^2}{\beta_{\parallel}}\right)^2 -x \right], \\	
f_{20} (x,y) &= \frac{s_{\parallel}^{y-2}}{\beta_{\parallel}^{y/2}} \left[ y (y-1) - 2(2y+1)\frac{s_{\parallel}^2}{\beta_{\parallel}} +4 \frac{s_{\parallel}^4}{\beta_{\parallel}^2}\right], \\	\end{align*}
together with the coefficients $P_{li}$, defined as
\begin{align}
P_{li} = \frac{(-1)^i}{2^l} \binom{l}{i} \binom{2l-2i}{l}.
\end{align}

\section{Laguerre Integrals}\label{appendix:lag_integrals}
We compute the Laguerre integrals $C_{pj}^m$ and $D_{pj}^m$ appearing  in \cref{coll_op_mom} by following two different approaches
The first approach is based on recursive relations between higher-order and lower-order integrals, while in the second approach the integrals are computed directly using hypergeometric functions.
In order to simplify the derivation in both approaches, we rewrite the integrals in \cref{lag-int1,lag-int2} using the fact that $s_{b \perp}^2 = \theta \beta_{\perp} s_{a\perp}^2 / \alpha_{\perp}$.
We then note that the integrals $C_{pj}^m$ and $D_{pj}^m$ in \cref{lag-int1,lag-int2} are only a function of $x=\theta \alpha_{\perp}^{-1}$ and $y=\theta  \alpha_{\perp}^{-1} \beta_{\perp}$, respectively, yielding
\begin{align}
C_{pj}^m (x) &\equiv \int_0^{\infty} x ^{m} z^{m} L_p (z)  L_j (z)  e^{-(1+x)z} \mr{d} z, \label{eq:laguerre_relabeled_C}\\
D_{pj}^m (y) &\equiv \int_0^{\infty} L_p (z) L_j (z) L_m (y z) e^{-(1+y )z} \mr{d} z. \label{eq:laguerre_relabeled_D}
\end{align} 

We first consider the approach based on recursive relations. We leverage the work in Refs. [\onlinecite{Gillis1962,Askey1977,Kleindienst1993,Khabibrakhmanov1998}], where closed analytical expressions for \cref{eq:laguerre_relabeled_C,eq:laguerre_relabeled_D} with $x=y=1$ are obtained.
We start by computing $C_{pj}^0 (x)$, performing the change of variables $z' = (1+x)z$ and using the transformation rule for Laguerre polynomials
\begin{equation}\label{transformation_lag}
L_n (x z) = \sum_{k=0}^n \binom{n}{n-k}x^k (1-x)^{n-k} L_k (z).
\end{equation}
From \cref{transformation_lag} it is straightforward to obtain that 
\begin{align}\label{cpj0}
C_{pj}^0 (x) = \frac{x^{p+j}}{(1+x)^{p+j+1}} \sum_{k=0}^{\mr{min}(p,j)} \binom{p}{p-k} \binom{j}{j-k} x^{-2k}. 
\end{align}
To calculate $C_{pj}^m$, for $m>0$, one can make use of the recursion relation for Laguerre polynomials in \cref{eq:laguerre1} to compute a recurrence formula between $C_{pj}^{m+1}$ and integrals of lower order in the index $m$.
We generalise the procedure outlined in Ref. [\onlinecite{Khabibrakhmanov1998}] for the case of $x=1$, to an arbitrary $x$. We thus have
\begin{align}
C^{m+1}_{pj} (x) =  (1+2p)x C^{m}_{pj} (x) - (p+1)x C^{m}_{p+1,j}(x) - p x C^{m}_{p-1,j} (x), \label{cpj_rec_formula}
\end{align}
with the boundary values of 
\begin{equation}
C^{m}_{p0} = C^{m}_{0p} = \sum_{k=0}^{\mr{min}(p,m)} \binom{p}{p-k} \binom{m}{m-k} \frac{m! (-1)^k x^{p+m-k}}{\left(1+x\right)^{p+m+1}}.
\end{equation}
%
%
%
To compute $D_{pj}^m$ for $m>0$, it is also possible to derive a recursion relation that involves integral of lower order in $m$. Using again \cref{eq:laguerre1}, we obtain
\begin{align}
(m+1)D_{pj}^{m+1}(y) = \left[ 2m+1-y(2p+1) \right] D_{pj}^m (y) \nonumber \\
- m D_{pj}^{m-1} (y) + y (p+1)D_{p+1,j}^{m} (y) + yp D_{p-1,j}^{m} (y), \label{dpj_rec_formula}
\end{align}
where 
\begin{align}
D_{pj}^0 (y) = C_{pj}^0 (y)
\end{align}
and with the boundary values of 
\begin{align}
D_{p0}^m (y) = D_{0p}^m (y) = \frac{y^p}{(1+y)^{p+m+1}} \frac{(p+m)!}{p!m!}. \label{d0}
\end{align}
%
  %
%
%
%
%
%

As a second approach, we note that the integrals $C^{m}_{pj}$ and $D_{pj}^{m}$ can also be obtained as a special case of the general expression for the integral of $k$ Laguerre polynomials \cite{Erdelyi1936}, i.e.
\begin{align}
&\int_0^{\infty} x^{\rho-1} e^{-\sigma x}  L_{n_1} (\lambda_1 x)  \ldots L_{n_k} (\lambda_k x) \mr{d}x \nonumber \\  & =\sigma^{-\rho} \Gamma(\rho) F_A^{(k)} \left( \rho , -n_1, \ldots, -n_k, \underbrace{1, \ldots, 1}_{k \, \mr{times}},  \frac{\lambda_1}{\sigma} , \ldots , \frac{\lambda_k}{\sigma} \right), \label{gen_lag} 
\end{align}
where $\rho, \sigma > 0 $ and $F_A^{(k)}$ is the first Lauricella hypergeometric function of $k$ variables, which is defined by
\begin{align}
&F_A^{(k)} \left( a, b_1, \ldots, b_k, c_1, \ldots, c_k, x_1,\ldots,x_k\right) \label{lauricella}\\
&= \sum_{m_1 = 0}^{\infty} \ldots \sum_{m_k = 0}^{\infty} \frac{(a)_{m_1+\ldots+m_k} (b_1)_{m_1} \times \ldots \times (b_k)_{m_k}}{(c_1)_{m_1} \times \ldots \times (c_k)_{m_k} m_1 ! \ldots m_k !}x_1^{m_1} \ldots x_k^{m_k}, \nonumber
\end{align}
where the Pochhammer symbol $(q)_n$ denotes the rising factorial:
\begin{equation}
    (q)_n = q(q+1)\ldots (q+n-1) = \frac{\Gamma(q+n)}{\Gamma(q)}.
\end{equation}
The $k=2$ Lauricella function is also known in the literature as the Appell hypergeometric function $F_2$ \cite{Gordon1929}.
It is worth mentioning that, although the Lauricella function is defined in general only for $\lvert x_1 \rvert + \ldots + \lvert x_k \rvert < 1$, in our specific case the integral is well defined and converges for any value of $m,p,j$ and $x >0$, since the arguments $b_1, \ldots, b_k$ are always negative and equal to $-1$, and therefore the sums in \cref{lauricella} are bounded.
Finally, leveraging the results of Ref. [\onlinecite{Erdelyi1936}], we write the integrals $C^{m}_{pj}$ and $D_{pj}^{m}$ as
\begin{align}
C_{pj}^m (x)   =  \frac{\Gamma(m+1) x^{m}}{(1+x)^{m+1}} F_2 \left( m+1, -p, -j, 1, 1, \frac{1}{1+x}, \frac{1}{1+x} \right) \label{cpj_appel},
\end{align}
and
\begin{equation}
D_{pj}^m (y) = \frac{1}{1+y }F_A^{(3)} \left( 1, -j, -m, -p, 1,1,1, \frac{1}{1+y}, \frac{y}{1+y}, \frac{1}{1+y}\right). \label{dpj_appel}
\end{equation}
It can be shown that \cref{cpj_appel} and \cref{dpj_appel} are equivalent to \cref{cpj_rec_formula} and \cref{dpj_rec_formula}, respectively, by verifying that they reduce to \cref{cpj0} when $m=0$, and that they satisfy the recursion relation in \cref{cpj_rec_formula,dpj_rec_formula} for $m>0$.

\bibliographystyle{apsrev4-2}
\bibliography{library}

\begin{thebibliography}{49}%
\makeatletter
\providecommand \@ifxundefined [1]{%
 \@ifx{#1\undefined}
}%
\providecommand \@ifnum [1]{%
 \ifnum #1\expandafter \@firstoftwo
 \else \expandafter \@secondoftwo
 \fi
}%
\providecommand \@ifx [1]{%
 \ifx #1\expandafter \@firstoftwo
 \else \expandafter \@secondoftwo
 \fi
}%
\providecommand \natexlab [1]{#1}%
\providecommand \enquote  [1]{``#1''}%
\providecommand \bibnamefont  [1]{#1}%
\providecommand \bibfnamefont [1]{#1}%
\providecommand \citenamefont [1]{#1}%
\providecommand \href@noop [0]{\@secondoftwo}%
\providecommand \href [0]{\begingroup \@sanitize@url \@href}%
\providecommand \@href[1]{\@@startlink{#1}\@@href}%
\providecommand \@@href[1]{\endgroup#1\@@endlink}%
\providecommand \@sanitize@url [0]{\catcode `\\12\catcode `\$12\catcode
  `\&12\catcode `\#12\catcode `\^12\catcode `\_12\catcode `\%12\relax}%
\providecommand \@@startlink[1]{}%
\providecommand \@@endlink[0]{}%
\providecommand \url  [0]{\begingroup\@sanitize@url \@url }%
\providecommand \@url [1]{\endgroup\@href {#1}{\urlprefix }}%
\providecommand \urlprefix  [0]{URL }%
\providecommand \Eprint [0]{\href }%
\providecommand \doibase [0]{https://doi.org/}%
\providecommand \selectlanguage [0]{\@gobble}%
\providecommand \bibinfo  [0]{\@secondoftwo}%
\providecommand \bibfield  [0]{\@secondoftwo}%
\providecommand \translation [1]{[#1]}%
\providecommand \BibitemOpen [0]{}%
\providecommand \bibitemStop [0]{}%
\providecommand \bibitemNoStop [0]{.\EOS\space}%
\providecommand \EOS [0]{\spacefactor3000\relax}%
\providecommand \BibitemShut  [1]{\csname bibitem#1\endcsname}%
\let\auto@bib@innerbib\@empty
\bibitem [{\citenamefont {Stangeby}(2000)}]{Stangeby2000}%
  \BibitemOpen
  \bibfield  {author} {\bibinfo {author} {\bibfnamefont {P.}~\bibnamefont
  {Stangeby}},\ }\href {https://doi.org/10.1201/9781420033328} {\emph {\bibinfo
  {title} {{The Plasma Boundary of Magnetic Fusion Devices}}}},\ Series in
  Plasma Physics\ (\bibinfo  {publisher} {CRC Press},\ \bibinfo {address} {Boca
  Raton},\ \bibinfo {year} {2000})\BibitemShut {NoStop}%
\bibitem [{\citenamefont {Ricci}(2015)}]{Ricci2015}%
  \BibitemOpen
  \bibfield  {author} {\bibinfo {author} {\bibfnamefont {P.}~\bibnamefont
  {Ricci}},\ }\href {https://doi.org/10.1017/S0022377814001202} {\bibfield
  {journal} {\bibinfo  {journal} {J. Plasma Phys.}\ }\textbf {\bibinfo {volume}
  {81}},\ \bibinfo {pages} {435810202} (\bibinfo {year} {2015})}\BibitemShut
  {NoStop}%
\bibitem [{\citenamefont {Theiler}\ \emph {et~al.}(2011)\citenamefont
  {Theiler}, \citenamefont {Furno}, \citenamefont {Fasoli}, \citenamefont
  {Ricci}, \citenamefont {Labit},\ and\ \citenamefont {Iraji}}]{Theiler2011}%
  \BibitemOpen
  \bibfield  {author} {\bibinfo {author} {\bibfnamefont {C.}~\bibnamefont
  {Theiler}}, \bibinfo {author} {\bibfnamefont {I.}~\bibnamefont {Furno}},
  \bibinfo {author} {\bibfnamefont {A.}~\bibnamefont {Fasoli}}, \bibinfo
  {author} {\bibfnamefont {P.}~\bibnamefont {Ricci}}, \bibinfo {author}
  {\bibfnamefont {B.}~\bibnamefont {Labit}},\ and\ \bibinfo {author}
  {\bibfnamefont {D.}~\bibnamefont {Iraji}},\ }\href
  {https://doi.org/10.1063/1.3562944} {\bibfield  {journal} {\bibinfo
  {journal} {Phys. Plasmas}\ }\textbf {\bibinfo {volume} {18}},\ \bibinfo
  {pages} {1} (\bibinfo {year} {2011})}\BibitemShut {NoStop}%
\bibitem [{\citenamefont {Jorge}\ \emph {et~al.}(2018)\citenamefont {Jorge},
  \citenamefont {Ricci},\ and\ \citenamefont {Loureiro}}]{Jorge2018}%
  \BibitemOpen
  \bibfield  {author} {\bibinfo {author} {\bibfnamefont {R.}~\bibnamefont
  {Jorge}}, \bibinfo {author} {\bibfnamefont {P.}~\bibnamefont {Ricci}},\ and\
  \bibinfo {author} {\bibfnamefont {N.~F.}\ \bibnamefont {Loureiro}},\ }\href
  {https://doi.org/10.1103/PhysRevLett.121.165001} {\bibfield  {journal}
  {\bibinfo  {journal} {Phys. Rev. Lett.}\ }\textbf {\bibinfo {volume} {121}},\
  \bibinfo {pages} {165001} (\bibinfo {year} {2018})}\BibitemShut {NoStop}%
\bibitem [{\citenamefont {Qin}\ \emph {et~al.}(2007)\citenamefont {Qin},
  \citenamefont {Cohen}, \citenamefont {Nevins},\ and\ \citenamefont
  {Xu}}]{Qin2007}%
  \BibitemOpen
  \bibfield  {author} {\bibinfo {author} {\bibfnamefont {H.}~\bibnamefont
  {Qin}}, \bibinfo {author} {\bibfnamefont {R.}~\bibnamefont {Cohen}}, \bibinfo
  {author} {\bibfnamefont {W.}~\bibnamefont {Nevins}},\ and\ \bibinfo {author}
  {\bibfnamefont {X.}~\bibnamefont {Xu}},\ }\href
  {https://doi.org/10.1063/1.2472596} {\bibfield  {journal} {\bibinfo
  {journal} {Phys. Plasmas}\ }\textbf {\bibinfo {volume} {14}},\ \bibinfo
  {pages} {056110} (\bibinfo {year} {2007})}\BibitemShut {NoStop}%
\bibitem [{\citenamefont {Hahm}\ \emph {et~al.}(2009)\citenamefont {Hahm},
  \citenamefont {Wang},\ and\ \citenamefont {Madsen}}]{Hahm2009}%
  \BibitemOpen
  \bibfield  {author} {\bibinfo {author} {\bibfnamefont {T.}~\bibnamefont
  {Hahm}}, \bibinfo {author} {\bibfnamefont {L.}~\bibnamefont {Wang}},\ and\
  \bibinfo {author} {\bibfnamefont {J.}~\bibnamefont {Madsen}},\ }\href
  {https://doi.org/10.1063/1.3073671} {\bibfield  {journal} {\bibinfo
  {journal} {Phys. Plasmas}\ }\textbf {\bibinfo {volume} {16}},\ \bibinfo
  {pages} {022305} (\bibinfo {year} {2009})}\BibitemShut {NoStop}%
\bibitem [{\citenamefont {Jorge}\ \emph
  {et~al.}(2019{\natexlab{a}})\citenamefont {Jorge}, \citenamefont {Ricci},
  \citenamefont {Brunner}, \citenamefont {Gamba}, \citenamefont {Konovets},
  \citenamefont {Loureiro}, \citenamefont {Perrone},\ and\ \citenamefont
  {Teixeira}}]{Jorge2019}%
  \BibitemOpen
  \bibfield  {author} {\bibinfo {author} {\bibfnamefont {R.}~\bibnamefont
  {Jorge}}, \bibinfo {author} {\bibfnamefont {P.}~\bibnamefont {Ricci}},
  \bibinfo {author} {\bibfnamefont {S.}~\bibnamefont {Brunner}}, \bibinfo
  {author} {\bibfnamefont {S.}~\bibnamefont {Gamba}}, \bibinfo {author}
  {\bibfnamefont {V.}~\bibnamefont {Konovets}}, \bibinfo {author}
  {\bibfnamefont {N.~F.}\ \bibnamefont {Loureiro}}, \bibinfo {author}
  {\bibfnamefont {L.~M.}\ \bibnamefont {Perrone}},\ and\ \bibinfo {author}
  {\bibfnamefont {N.}~\bibnamefont {Teixeira}},\ }\href@noop {} {\bibfield
  {journal} {\bibinfo  {journal} {J. Plasma Phys.}\ }\textbf {\bibinfo {volume}
  {85}},\ \bibinfo {pages} {905850211} (\bibinfo {year}
  {2019}{\natexlab{a}})}\BibitemShut {NoStop}%
\bibitem [{\citenamefont {Frei}\ \emph {et~al.}(2020)\citenamefont {Frei},
  \citenamefont {Jorge},\ and\ \citenamefont {Ricci}}]{Frei2020}%
  \BibitemOpen
  \bibfield  {author} {\bibinfo {author} {\bibfnamefont {B.~J.}\ \bibnamefont
  {Frei}}, \bibinfo {author} {\bibfnamefont {R.}~\bibnamefont {Jorge}},\ and\
  \bibinfo {author} {\bibfnamefont {P.}~\bibnamefont {Ricci}},\ }\href
  {https://doi.org/10.1017/S0022377820000100} {\bibfield  {journal} {\bibinfo
  {journal} {J. Plasma Phys.}\ }\textbf {\bibinfo {volume} {86}},\ \bibinfo
  {pages} {905860205} (\bibinfo {year} {2020})}\BibitemShut {NoStop}%
\bibitem [{\citenamefont {Chang}\ \emph {et~al.}(2009)\citenamefont {Chang},
  \citenamefont {Ku}, \citenamefont {Diamond}, \citenamefont {Lin},
  \citenamefont {Parker}, \citenamefont {Hahm},\ and\ \citenamefont
  {Samatova}}]{Chang2009a}%
  \BibitemOpen
  \bibfield  {author} {\bibinfo {author} {\bibfnamefont {C.~S.}\ \bibnamefont
  {Chang}}, \bibinfo {author} {\bibfnamefont {S.}~\bibnamefont {Ku}}, \bibinfo
  {author} {\bibfnamefont {P.~H.}\ \bibnamefont {Diamond}}, \bibinfo {author}
  {\bibfnamefont {Z.}~\bibnamefont {Lin}}, \bibinfo {author} {\bibfnamefont
  {S.}~\bibnamefont {Parker}}, \bibinfo {author} {\bibfnamefont {T.~S.}\
  \bibnamefont {Hahm}},\ and\ \bibinfo {author} {\bibfnamefont
  {N.}~\bibnamefont {Samatova}},\ }\href {https://doi.org/10.1063/1.3099329}
  {\bibfield  {journal} {\bibinfo  {journal} {Phys. Plasmas}\ }\textbf
  {\bibinfo {volume} {16}},\ \bibinfo {pages} {056108} (\bibinfo {year}
  {2009})}\BibitemShut {NoStop}%
\bibitem [{\citenamefont {Shi}\ \emph {et~al.}(2017)\citenamefont {Shi},
  \citenamefont {Hammett}, \citenamefont {Stoltzfus-Dueck},\ and\ \citenamefont
  {Hakim}}]{Shi2017}%
  \BibitemOpen
  \bibfield  {author} {\bibinfo {author} {\bibfnamefont {E.~L.}\ \bibnamefont
  {Shi}}, \bibinfo {author} {\bibfnamefont {G.~W.}\ \bibnamefont {Hammett}},
  \bibinfo {author} {\bibfnamefont {T.}~\bibnamefont {Stoltzfus-Dueck}},\ and\
  \bibinfo {author} {\bibfnamefont {A.}~\bibnamefont {Hakim}},\ }\href
  {https://doi.org/10.1017/S002237781700037X} {\bibfield  {journal} {\bibinfo
  {journal} {J. Plasma Phys.}\ }\textbf {\bibinfo {volume} {83}},\ \bibinfo
  {pages} {905830304} (\bibinfo {year} {2017})}\BibitemShut {NoStop}%
\bibitem [{\citenamefont {Pan}\ \emph {et~al.}(2018)\citenamefont {Pan},
  \citenamefont {Told}, \citenamefont {Shi}, \citenamefont {Hammett},\ and\
  \citenamefont {Jenko}}]{Pan2018}%
  \BibitemOpen
  \bibfield  {author} {\bibinfo {author} {\bibfnamefont {Q.}~\bibnamefont
  {Pan}}, \bibinfo {author} {\bibfnamefont {D.}~\bibnamefont {Told}}, \bibinfo
  {author} {\bibfnamefont {E.}~\bibnamefont {Shi}}, \bibinfo {author}
  {\bibfnamefont {G.~W.}\ \bibnamefont {Hammett}},\ and\ \bibinfo {author}
  {\bibfnamefont {F.}~\bibnamefont {Jenko}},\ }\href
  {https://doi.org/10.1063/1.5008895} {\bibfield  {journal} {\bibinfo
  {journal} {Phys. Plasmas}\ }\textbf {\bibinfo {volume} {25}},\ \bibinfo
  {pages} {062303} (\bibinfo {year} {2018})}\BibitemShut {NoStop}%
\bibitem [{\citenamefont {Ricci}\ \emph {et~al.}(2012)\citenamefont {Ricci},
  \citenamefont {Halpern}, \citenamefont {Jolliet}, \citenamefont {Loizu},
  \citenamefont {Mosetto}, \citenamefont {Fasoli}, \citenamefont {Furno},\ and\
  \citenamefont {Theiler}}]{Ricci2012a}%
  \BibitemOpen
  \bibfield  {author} {\bibinfo {author} {\bibfnamefont {P.}~\bibnamefont
  {Ricci}}, \bibinfo {author} {\bibfnamefont {F.~D.}\ \bibnamefont {Halpern}},
  \bibinfo {author} {\bibfnamefont {S.}~\bibnamefont {Jolliet}}, \bibinfo
  {author} {\bibfnamefont {J.}~\bibnamefont {Loizu}}, \bibinfo {author}
  {\bibfnamefont {A.}~\bibnamefont {Mosetto}}, \bibinfo {author} {\bibfnamefont
  {A.}~\bibnamefont {Fasoli}}, \bibinfo {author} {\bibfnamefont
  {I.}~\bibnamefont {Furno}},\ and\ \bibinfo {author} {\bibfnamefont
  {C.}~\bibnamefont {Theiler}},\ }\href
  {https://doi.org/10.1088/0741-3335/54/12/124047} {\bibfield  {journal}
  {\bibinfo  {journal} {Plasma Phys. Control. Fusion}\ }\textbf {\bibinfo
  {volume} {54}},\ \bibinfo {pages} {124047} (\bibinfo {year}
  {2012})}\BibitemShut {NoStop}%
\bibitem [{\citenamefont {Dudson}\ \emph {et~al.}(2009)\citenamefont {Dudson},
  \citenamefont {Umansky}, \citenamefont {Xu}, \citenamefont {Snyder},\ and\
  \citenamefont {Wilson}}]{Dudson2009}%
  \BibitemOpen
  \bibfield  {author} {\bibinfo {author} {\bibfnamefont {B.}~\bibnamefont
  {Dudson}}, \bibinfo {author} {\bibfnamefont {M.}~\bibnamefont {Umansky}},
  \bibinfo {author} {\bibfnamefont {X.}~\bibnamefont {Xu}}, \bibinfo {author}
  {\bibfnamefont {P.}~\bibnamefont {Snyder}},\ and\ \bibinfo {author}
  {\bibfnamefont {H.}~\bibnamefont {Wilson}},\ }\href
  {https://doi.org/10.1016/j.cpc.2009.03.008} {\bibfield  {journal} {\bibinfo
  {journal} {Comput. Phys. Commun.}\ }\textbf {\bibinfo {volume} {180}},\
  \bibinfo {pages} {1467} (\bibinfo {year} {2009})}\BibitemShut {NoStop}%
\bibitem [{\citenamefont {Tamain}\ \emph {et~al.}(2009)\citenamefont {Tamain},
  \citenamefont {Ghendrih}, \citenamefont {Tsitrone}, \citenamefont {Sarazin},
  \citenamefont {Garbet}, \citenamefont {Grandgirard}, \citenamefont {Gunn},
  \citenamefont {Serre}, \citenamefont {Ciraolo},\ and\ \citenamefont
  {Chiavassa}}]{Tamain2009}%
  \BibitemOpen
  \bibfield  {author} {\bibinfo {author} {\bibfnamefont {P.}~\bibnamefont
  {Tamain}}, \bibinfo {author} {\bibfnamefont {P.}~\bibnamefont {Ghendrih}},
  \bibinfo {author} {\bibfnamefont {E.}~\bibnamefont {Tsitrone}}, \bibinfo
  {author} {\bibfnamefont {Y.}~\bibnamefont {Sarazin}}, \bibinfo {author}
  {\bibfnamefont {X.}~\bibnamefont {Garbet}}, \bibinfo {author} {\bibfnamefont
  {V.}~\bibnamefont {Grandgirard}}, \bibinfo {author} {\bibfnamefont
  {J.}~\bibnamefont {Gunn}}, \bibinfo {author} {\bibfnamefont {E.}~\bibnamefont
  {Serre}}, \bibinfo {author} {\bibfnamefont {G.}~\bibnamefont {Ciraolo}},\
  and\ \bibinfo {author} {\bibfnamefont {G.}~\bibnamefont {Chiavassa}},\ }\href
  {https://doi.org/10.1016/j.jnucmat.2009.01.062} {\bibfield  {journal}
  {\bibinfo  {journal} {J. Nucl. Mater.}\ }\textbf {\bibinfo {volume} {390}},\
  \bibinfo {pages} {347} (\bibinfo {year} {2009})}\BibitemShut {NoStop}%
\bibitem [{\citenamefont {Easy}\ \emph {et~al.}(2014)\citenamefont {Easy},
  \citenamefont {Militello}, \citenamefont {Omotani}, \citenamefont {Dudson},
  \citenamefont {Havlkov}, \citenamefont {Tamain}, \citenamefont {Naulin},\
  and\ \citenamefont {Nielsen}}]{Easy2014}%
  \BibitemOpen
  \bibfield  {author} {\bibinfo {author} {\bibfnamefont {L.}~\bibnamefont
  {Easy}}, \bibinfo {author} {\bibfnamefont {F.}~\bibnamefont {Militello}},
  \bibinfo {author} {\bibfnamefont {J.}~\bibnamefont {Omotani}}, \bibinfo
  {author} {\bibfnamefont {B.}~\bibnamefont {Dudson}}, \bibinfo {author}
  {\bibfnamefont {E.}~\bibnamefont {Havlkov}}, \bibinfo {author} {\bibfnamefont
  {P.}~\bibnamefont {Tamain}}, \bibinfo {author} {\bibfnamefont
  {V.}~\bibnamefont {Naulin}},\ and\ \bibinfo {author} {\bibfnamefont
  {A.}~\bibnamefont {Nielsen}},\ }\href {https://doi.org/10.1063/1.4904207}
  {\bibfield  {journal} {\bibinfo  {journal} {Phys. Plasmas}\ }\textbf
  {\bibinfo {volume} {21}},\ \bibinfo {pages} {122515} (\bibinfo {year}
  {2014})}\BibitemShut {NoStop}%
\bibitem [{\citenamefont {Halpern}\ \emph {et~al.}(2016)\citenamefont
  {Halpern}, \citenamefont {Ricci}, \citenamefont {Jolliet}, \citenamefont
  {Loizu}, \citenamefont {Morales}, \citenamefont {Mosetto}, \citenamefont
  {Musil}, \citenamefont {Riva}, \citenamefont {Tran},\ and\ \citenamefont
  {Wersal}}]{Halpern2016a}%
  \BibitemOpen
  \bibfield  {author} {\bibinfo {author} {\bibfnamefont {F.}~\bibnamefont
  {Halpern}}, \bibinfo {author} {\bibfnamefont {P.}~\bibnamefont {Ricci}},
  \bibinfo {author} {\bibfnamefont {S.}~\bibnamefont {Jolliet}}, \bibinfo
  {author} {\bibfnamefont {J.}~\bibnamefont {Loizu}}, \bibinfo {author}
  {\bibfnamefont {J.}~\bibnamefont {Morales}}, \bibinfo {author} {\bibfnamefont
  {A.}~\bibnamefont {Mosetto}}, \bibinfo {author} {\bibfnamefont
  {F.}~\bibnamefont {Musil}}, \bibinfo {author} {\bibfnamefont
  {F.}~\bibnamefont {Riva}}, \bibinfo {author} {\bibfnamefont {T.}~\bibnamefont
  {Tran}},\ and\ \bibinfo {author} {\bibfnamefont {C.}~\bibnamefont {Wersal}},\
  }\href {https://doi.org/10.1016/j.jcp.2016.03.040} {\bibfield  {journal}
  {\bibinfo  {journal} {J. Comput. Phys.}\ }\textbf {\bibinfo {volume} {315}},\
  \bibinfo {pages} {388} (\bibinfo {year} {2016})}\BibitemShut {NoStop}%
\bibitem [{\citenamefont {Madsen}\ \emph {et~al.}(2016)\citenamefont {Madsen},
  \citenamefont {Naulin}, \citenamefont {Nielsen},\ and\ \citenamefont
  {Rasmussen}}]{Madsen2016}%
  \BibitemOpen
  \bibfield  {author} {\bibinfo {author} {\bibfnamefont {J.}~\bibnamefont
  {Madsen}}, \bibinfo {author} {\bibfnamefont {V.}~\bibnamefont {Naulin}},
  \bibinfo {author} {\bibfnamefont {A.}~\bibnamefont {Nielsen}},\ and\ \bibinfo
  {author} {\bibfnamefont {J.}~\bibnamefont {Rasmussen}},\ }\href
  {https://doi.org/10.1063/1.4943199} {\bibfield  {journal} {\bibinfo
  {journal} {Phys. Plasmas}\ }\textbf {\bibinfo {volume} {23}},\ \bibinfo
  {pages} {032306} (\bibinfo {year} {2016})}\BibitemShut {NoStop}%
\bibitem [{\citenamefont {Zhu}\ \emph {et~al.}(2017)\citenamefont {Zhu},
  \citenamefont {Francisquez},\ and\ \citenamefont {Rogers}}]{Zhu2017}%
  \BibitemOpen
  \bibfield  {author} {\bibinfo {author} {\bibfnamefont {B.}~\bibnamefont
  {Zhu}}, \bibinfo {author} {\bibfnamefont {M.}~\bibnamefont {Francisquez}},\
  and\ \bibinfo {author} {\bibfnamefont {B.~N.}\ \bibnamefont {Rogers}},\
  }\href {https://doi.org/10.1063/1.4978885} {\bibfield  {journal} {\bibinfo
  {journal} {Phys. Plasmas}\ }\textbf {\bibinfo {volume} {24}},\ \bibinfo
  {pages} {055903} (\bibinfo {year} {2017})}\BibitemShut {NoStop}%
\bibitem [{\citenamefont {Paruta}\ \emph {et~al.}(2018)\citenamefont {Paruta},
  \citenamefont {Ricci}, \citenamefont {Riva}, \citenamefont {Wersal},
  \citenamefont {Beadle},\ and\ \citenamefont {Frei}}]{Paruta2018}%
  \BibitemOpen
  \bibfield  {author} {\bibinfo {author} {\bibfnamefont {P.}~\bibnamefont
  {Paruta}}, \bibinfo {author} {\bibfnamefont {P.}~\bibnamefont {Ricci}},
  \bibinfo {author} {\bibfnamefont {F.}~\bibnamefont {Riva}}, \bibinfo {author}
  {\bibfnamefont {C.}~\bibnamefont {Wersal}}, \bibinfo {author} {\bibfnamefont
  {C.}~\bibnamefont {Beadle}},\ and\ \bibinfo {author} {\bibfnamefont
  {B.}~\bibnamefont {Frei}},\ }\href {https://doi.org/10.1063/1.5047741}
  {\bibfield  {journal} {\bibinfo  {journal} {Phys. Plasmas}\ }\textbf
  {\bibinfo {volume} {25}},\ \bibinfo {pages} {112301} (\bibinfo {year}
  {2018})}\BibitemShut {NoStop}%
\bibitem [{\citenamefont {Zeiler}\ \emph {et~al.}(1997)\citenamefont {Zeiler},
  \citenamefont {Drake},\ and\ \citenamefont {Rogers}}]{Zeiler1997}%
  \BibitemOpen
  \bibfield  {author} {\bibinfo {author} {\bibfnamefont {A.}~\bibnamefont
  {Zeiler}}, \bibinfo {author} {\bibfnamefont {J.}~\bibnamefont {Drake}},\ and\
  \bibinfo {author} {\bibfnamefont {B.}~\bibnamefont {Rogers}},\ }\href
  {https://doi.org/10.1063/1.872368} {\bibfield  {journal} {\bibinfo  {journal}
  {Phys. Plasmas}\ }\textbf {\bibinfo {volume} {4}},\ \bibinfo {pages} {2134}
  (\bibinfo {year} {1997})}\BibitemShut {NoStop}%
\bibitem [{\citenamefont {Scott}(1997)}]{Scott1997}%
  \BibitemOpen
  \bibfield  {author} {\bibinfo {author} {\bibfnamefont {B.}~\bibnamefont
  {Scott}},\ }\href {https://doi.org/10.1088/0741-3335/39/10/010} {\bibfield
  {journal} {\bibinfo  {journal} {Plasma Phys. Control. Fusion}\ }\textbf
  {\bibinfo {volume} {39}},\ \bibinfo {pages} {1635} (\bibinfo {year}
  {1997})}\BibitemShut {NoStop}%
\bibitem [{\citenamefont {Madsen}(2013)}]{Madsen2013}%
  \BibitemOpen
  \bibfield  {author} {\bibinfo {author} {\bibfnamefont {J.}~\bibnamefont
  {Madsen}},\ }\href {https://doi.org/10.1063/1.4813241} {\bibfield  {journal}
  {\bibinfo  {journal} {Phys. Plasmas}\ }\textbf {\bibinfo {volume} {20}},\
  \bibinfo {pages} {072301} (\bibinfo {year} {2013})}\BibitemShut {NoStop}%
\bibitem [{\citenamefont {Lonnroth}\ \emph {et~al.}(2006)\citenamefont
  {Lonnroth}, \citenamefont {Bateman}, \citenamefont {B{\'{e}}coulet},
  \citenamefont {Beyer}, \citenamefont {Corrigan}, \citenamefont {Figarella},
  \citenamefont {Fundamenski}, \citenamefont {Garcia}, \citenamefont {Garbet},
  \citenamefont {Huysmans}, \citenamefont {Janeschitz}, \citenamefont
  {Johnson}, \citenamefont {Kiviniemi}, \citenamefont {Kuhn}, \citenamefont
  {Kritz}, \citenamefont {Loarte}, \citenamefont {Naulin}, \citenamefont
  {Nave}, \citenamefont {Onjun}, \citenamefont {Pacher}, \citenamefont
  {Pacher}, \citenamefont {Pankin}, \citenamefont {Parail}, \citenamefont
  {Pitts}, \citenamefont {Saibene}, \citenamefont {Snyder}, \citenamefont
  {Spence}, \citenamefont {Tskhakaya},\ and\ \citenamefont
  {Wilson}}]{Lonnroth2006}%
  \BibitemOpen
  \bibfield  {author} {\bibinfo {author} {\bibfnamefont {J.~S.}\ \bibnamefont
  {Lonnroth}}, \bibinfo {author} {\bibfnamefont {G.}~\bibnamefont {Bateman}},
  \bibinfo {author} {\bibfnamefont {M.}~\bibnamefont {B{\'{e}}coulet}},
  \bibinfo {author} {\bibfnamefont {P.}~\bibnamefont {Beyer}}, \bibinfo
  {author} {\bibfnamefont {G.}~\bibnamefont {Corrigan}}, \bibinfo {author}
  {\bibfnamefont {C.}~\bibnamefont {Figarella}}, \bibinfo {author}
  {\bibfnamefont {W.}~\bibnamefont {Fundamenski}}, \bibinfo {author}
  {\bibfnamefont {O.~E.}\ \bibnamefont {Garcia}}, \bibinfo {author}
  {\bibfnamefont {X.}~\bibnamefont {Garbet}}, \bibinfo {author} {\bibfnamefont
  {G.}~\bibnamefont {Huysmans}}, \bibinfo {author} {\bibfnamefont
  {G.}~\bibnamefont {Janeschitz}}, \bibinfo {author} {\bibfnamefont
  {T.}~\bibnamefont {Johnson}}, \bibinfo {author} {\bibfnamefont
  {T.}~\bibnamefont {Kiviniemi}}, \bibinfo {author} {\bibfnamefont
  {S.}~\bibnamefont {Kuhn}}, \bibinfo {author} {\bibfnamefont {A.}~\bibnamefont
  {Kritz}}, \bibinfo {author} {\bibfnamefont {A.}~\bibnamefont {Loarte}},
  \bibinfo {author} {\bibfnamefont {V.}~\bibnamefont {Naulin}}, \bibinfo
  {author} {\bibfnamefont {F.}~\bibnamefont {Nave}}, \bibinfo {author}
  {\bibfnamefont {T.}~\bibnamefont {Onjun}}, \bibinfo {author} {\bibfnamefont
  {G.~W.}\ \bibnamefont {Pacher}}, \bibinfo {author} {\bibfnamefont {H.~D.}\
  \bibnamefont {Pacher}}, \bibinfo {author} {\bibfnamefont {A.}~\bibnamefont
  {Pankin}}, \bibinfo {author} {\bibfnamefont {V.}~\bibnamefont {Parail}},
  \bibinfo {author} {\bibfnamefont {R.}~\bibnamefont {Pitts}}, \bibinfo
  {author} {\bibfnamefont {G.}~\bibnamefont {Saibene}}, \bibinfo {author}
  {\bibfnamefont {P.}~\bibnamefont {Snyder}}, \bibinfo {author} {\bibfnamefont
  {J.}~\bibnamefont {Spence}}, \bibinfo {author} {\bibfnamefont
  {D.}~\bibnamefont {Tskhakaya}},\ and\ \bibinfo {author} {\bibfnamefont
  {H.}~\bibnamefont {Wilson}},\ }\href {https://doi.org/10.1002/ctpp.200610070}
  {\bibfield  {journal} {\bibinfo  {journal} {Contrib. to Plasma Phys.}\
  }\textbf {\bibinfo {volume} {46}},\ \bibinfo {pages} {726} (\bibinfo {year}
  {2006})}\BibitemShut {NoStop}%
\bibitem [{\citenamefont {Leonard}(2014)}]{Leonard2014}%
  \BibitemOpen
  \bibfield  {author} {\bibinfo {author} {\bibfnamefont {A.}~\bibnamefont
  {Leonard}},\ }\href {https://doi.org/10.1063/1.4894742} {\bibfield  {journal}
  {\bibinfo  {journal} {Phys. Plasmas}\ }\textbf {\bibinfo {volume} {21}},\
  \bibinfo {pages} {090501} (\bibinfo {year} {2014})}\BibitemShut {NoStop}%
\bibitem [{\citenamefont {Endler}\ \emph {et~al.}(1995)\citenamefont {Endler},
  \citenamefont {Niedermeyer}, \citenamefont {Giannone}, \citenamefont
  {Kolzhauer}, \citenamefont {Rudyj}, \citenamefont {Theimer},\ and\
  \citenamefont {Tsois}}]{Endler1995}%
  \BibitemOpen
  \bibfield  {author} {\bibinfo {author} {\bibfnamefont {M.}~\bibnamefont
  {Endler}}, \bibinfo {author} {\bibfnamefont {H.}~\bibnamefont {Niedermeyer}},
  \bibinfo {author} {\bibfnamefont {L.}~\bibnamefont {Giannone}}, \bibinfo
  {author} {\bibfnamefont {E.}~\bibnamefont {Kolzhauer}}, \bibinfo {author}
  {\bibfnamefont {A.}~\bibnamefont {Rudyj}}, \bibinfo {author} {\bibfnamefont
  {G.}~\bibnamefont {Theimer}},\ and\ \bibinfo {author} {\bibfnamefont
  {N.}~\bibnamefont {Tsois}},\ }\href
  {https://doi.org/10.1088/0029-5515/35/11/I01} {\bibfield  {journal} {\bibinfo
   {journal} {Nucl. Fusion}\ }\textbf {\bibinfo {volume} {35}},\ \bibinfo
  {pages} {1307} (\bibinfo {year} {1995})}\BibitemShut {NoStop}%
\bibitem [{\citenamefont {Agostini}\ \emph {et~al.}(2011)\citenamefont
  {Agostini}, \citenamefont {Terry}, \citenamefont {Scarin},\ and\
  \citenamefont {Zweben}}]{Agostini2011}%
  \BibitemOpen
  \bibfield  {author} {\bibinfo {author} {\bibfnamefont {M.}~\bibnamefont
  {Agostini}}, \bibinfo {author} {\bibfnamefont {J.}~\bibnamefont {Terry}},
  \bibinfo {author} {\bibfnamefont {P.}~\bibnamefont {Scarin}},\ and\ \bibinfo
  {author} {\bibfnamefont {S.}~\bibnamefont {Zweben}},\ }\href
  {https://doi.org/10.1088/0029-5515/51/5/053020} {\bibfield  {journal}
  {\bibinfo  {journal} {Nucl. Fusion}\ }\textbf {\bibinfo {volume} {51}},\
  \bibinfo {pages} {053020} (\bibinfo {year} {2011})}\BibitemShut {NoStop}%
\bibitem [{\citenamefont {Carralero}\ \emph {et~al.}(2014)\citenamefont
  {Carralero}, \citenamefont {Birkenmeier}, \citenamefont {M{\"{u}}ller},
  \citenamefont {Manz}, \citenamefont {DeMarne}, \citenamefont {M{\"{u}}ller},
  \citenamefont {Reimold}, \citenamefont {Stroth}, \citenamefont {Wischmeier},\
  and\ \citenamefont {Wolfrum}}]{Carralero2014}%
  \BibitemOpen
  \bibfield  {author} {\bibinfo {author} {\bibfnamefont {D.}~\bibnamefont
  {Carralero}}, \bibinfo {author} {\bibfnamefont {G.}~\bibnamefont
  {Birkenmeier}}, \bibinfo {author} {\bibfnamefont {H.}~\bibnamefont
  {M{\"{u}}ller}}, \bibinfo {author} {\bibfnamefont {P.}~\bibnamefont {Manz}},
  \bibinfo {author} {\bibfnamefont {P.}~\bibnamefont {DeMarne}}, \bibinfo
  {author} {\bibfnamefont {S.}~\bibnamefont {M{\"{u}}ller}}, \bibinfo {author}
  {\bibfnamefont {F.}~\bibnamefont {Reimold}}, \bibinfo {author} {\bibfnamefont
  {U.}~\bibnamefont {Stroth}}, \bibinfo {author} {\bibfnamefont
  {M.}~\bibnamefont {Wischmeier}},\ and\ \bibinfo {author} {\bibfnamefont
  {E.}~\bibnamefont {Wolfrum}},\ }\href
  {https://doi.org/10.1088/0029-5515/54/12/123005} {\bibfield  {journal}
  {\bibinfo  {journal} {Nucl. Fusion}\ }\textbf {\bibinfo {volume} {54}},\
  \bibinfo {pages} {123005} (\bibinfo {year} {2014})}\BibitemShut {NoStop}%
\bibitem [{\citenamefont {Jorge}\ \emph {et~al.}(2017)\citenamefont {Jorge},
  \citenamefont {Ricci},\ and\ \citenamefont {Loureiro}}]{Jorge2017}%
  \BibitemOpen
  \bibfield  {author} {\bibinfo {author} {\bibfnamefont {R.}~\bibnamefont
  {Jorge}}, \bibinfo {author} {\bibfnamefont {P.}~\bibnamefont {Ricci}},\ and\
  \bibinfo {author} {\bibfnamefont {N.~F.}\ \bibnamefont {Loureiro}},\ }\href
  {https://doi.org/10.1017/S002237781700085X} {\bibfield  {journal} {\bibinfo
  {journal} {J. Plasma Phys.}\ }\textbf {\bibinfo {volume} {83}},\ \bibinfo
  {pages} {905830606} (\bibinfo {year} {2017})}\BibitemShut {NoStop}%
\bibitem [{\citenamefont {Mandell}\ \emph {et~al.}(2018)\citenamefont
  {Mandell}, \citenamefont {Dorland},\ and\ \citenamefont
  {Landreman}}]{Mandell2018}%
  \BibitemOpen
  \bibfield  {author} {\bibinfo {author} {\bibfnamefont {N.~R.}\ \bibnamefont
  {Mandell}}, \bibinfo {author} {\bibfnamefont {W.}~\bibnamefont {Dorland}},\
  and\ \bibinfo {author} {\bibfnamefont {M.}~\bibnamefont {Landreman}},\ }\href
  {https://doi.org/10.1017/S0022377818000041} {\bibfield  {journal} {\bibinfo
  {journal} {J. Plasma Phys.}\ }\textbf {\bibinfo {volume} {84}},\ \bibinfo
  {pages} {905840108} (\bibinfo {year} {2018})}\BibitemShut {NoStop}%
\bibitem [{\citenamefont {Jorge}\ \emph
  {et~al.}(2019{\natexlab{b}})\citenamefont {Jorge}, \citenamefont {Frei},\
  and\ \citenamefont {Ricci}}]{Jorge2019a}%
  \BibitemOpen
  \bibfield  {author} {\bibinfo {author} {\bibfnamefont {R.}~\bibnamefont
  {Jorge}}, \bibinfo {author} {\bibfnamefont {B.~J.}\ \bibnamefont {Frei}},\
  and\ \bibinfo {author} {\bibfnamefont {P.}~\bibnamefont {Ricci}},\ }\href
  {https://doi.org/10.1017/S0022377819000734} {\bibfield  {journal} {\bibinfo
  {journal} {J. Plasma Phys.}\ }\textbf {\bibinfo {volume} {85}},\ \bibinfo
  {pages} {905850604} (\bibinfo {year} {2019}{\natexlab{b}})}\BibitemShut
  {NoStop}%
\bibitem [{\citenamefont {Loizu}\ \emph {et~al.}(2011)\citenamefont {Loizu},
  \citenamefont {Ricci},\ and\ \citenamefont {Theiler}}]{Loizu2011}%
  \BibitemOpen
  \bibfield  {author} {\bibinfo {author} {\bibfnamefont {J.}~\bibnamefont
  {Loizu}}, \bibinfo {author} {\bibfnamefont {P.}~\bibnamefont {Ricci}},\ and\
  \bibinfo {author} {\bibfnamefont {C.}~\bibnamefont {Theiler}},\ }\href
  {https://doi.org/10.1103/PhysRevE.83.016406} {\bibfield  {journal} {\bibinfo
  {journal} {Phys. Rev. E - Stat. Nonlinear, Soft Matter Phys.}\ }\textbf
  {\bibinfo {volume} {83}},\ \bibinfo {pages} {016406} (\bibinfo {year}
  {2011})}\BibitemShut {NoStop}%
\bibitem [{\citenamefont {Omotani}\ \emph {et~al.}(2015)\citenamefont
  {Omotani}, \citenamefont {Dudson}, \citenamefont {Havl{\'{i}}ckova},\ and\
  \citenamefont {Umansky}}]{Omotani2015}%
  \BibitemOpen
  \bibfield  {author} {\bibinfo {author} {\bibfnamefont {J.}~\bibnamefont
  {Omotani}}, \bibinfo {author} {\bibfnamefont {B.}~\bibnamefont {Dudson}},
  \bibinfo {author} {\bibfnamefont {E.}~\bibnamefont {Havl{\'{i}}ckova}},\ and\
  \bibinfo {author} {\bibfnamefont {M.}~\bibnamefont {Umansky}},\ }\href
  {https://doi.org/10.1016/j.jnucmat.2014.10.040} {\bibfield  {journal}
  {\bibinfo  {journal} {J. Nucl. Mater.}\ }\textbf {\bibinfo {volume} {463}},\
  \bibinfo {pages} {769} (\bibinfo {year} {2015})}\BibitemShut {NoStop}%
\bibitem [{\citenamefont {Geraldini}\ \emph {et~al.}(2018)\citenamefont
  {Geraldini}, \citenamefont {Parra},\ and\ \citenamefont
  {Militello}}]{Geraldini2018}%
  \BibitemOpen
  \bibfield  {author} {\bibinfo {author} {\bibfnamefont {A.}~\bibnamefont
  {Geraldini}}, \bibinfo {author} {\bibfnamefont {F.~I.}\ \bibnamefont
  {Parra}},\ and\ \bibinfo {author} {\bibfnamefont {F.}~\bibnamefont
  {Militello}},\ }\href {https://doi.org/10.1088/1361-6587/aae29f} {\bibfield
  {journal} {\bibinfo  {journal} {Plasma Phys. Control. Fusion}\ }\textbf
  {\bibinfo {volume} {60}},\ \bibinfo {pages} {125002} (\bibinfo {year}
  {2018})}\BibitemShut {NoStop}%
\bibitem [{\citenamefont {Mosetto}\ \emph {et~al.}(2015)\citenamefont
  {Mosetto}, \citenamefont {Halpern}, \citenamefont {Jolliet}, \citenamefont
  {Loizu},\ and\ \citenamefont {Ricci}}]{Mosetto2015}%
  \BibitemOpen
  \bibfield  {author} {\bibinfo {author} {\bibfnamefont {A.}~\bibnamefont
  {Mosetto}}, \bibinfo {author} {\bibfnamefont {F.}~\bibnamefont {Halpern}},
  \bibinfo {author} {\bibfnamefont {S.}~\bibnamefont {Jolliet}}, \bibinfo
  {author} {\bibfnamefont {J.}~\bibnamefont {Loizu}},\ and\ \bibinfo {author}
  {\bibfnamefont {P.}~\bibnamefont {Ricci}},\ }\href
  {https://doi.org/10.1063/1.4904300} {\bibfield  {journal} {\bibinfo
  {journal} {Phys. Plasmas}\ }\textbf {\bibinfo {volume} {22}},\ \bibinfo
  {pages} {012308} (\bibinfo {year} {2015})}\BibitemShut {NoStop}%
\bibitem [{\citenamefont {Rosenbluth}\ \emph {et~al.}(1957)\citenamefont
  {Rosenbluth}, \citenamefont {MacDonald},\ and\ \citenamefont
  {Judd}}]{Rosenbluth1957}%
  \BibitemOpen
  \bibfield  {author} {\bibinfo {author} {\bibfnamefont {M.~N.}\ \bibnamefont
  {Rosenbluth}}, \bibinfo {author} {\bibfnamefont {W.~M.}\ \bibnamefont
  {MacDonald}},\ and\ \bibinfo {author} {\bibfnamefont {D.~L.}\ \bibnamefont
  {Judd}},\ }\href {https://doi.org/10.1103/PhysRev.107.1} {\bibfield
  {journal} {\bibinfo  {journal} {Phys. Rev.}\ }\textbf {\bibinfo {volume}
  {107}},\ \bibinfo {pages} {1} (\bibinfo {year} {1957})}\BibitemShut {NoStop}%
\bibitem [{\citenamefont {Hakim}\ \emph {et~al.}(2020)\citenamefont {Hakim},
  \citenamefont {Francisquez}, \citenamefont {Juno},\ and\ \citenamefont
  {Hammett}}]{Hakim2020}%
  \BibitemOpen
  \bibfield  {author} {\bibinfo {author} {\bibfnamefont {A.}~\bibnamefont
  {Hakim}}, \bibinfo {author} {\bibfnamefont {M.}~\bibnamefont {Francisquez}},
  \bibinfo {author} {\bibfnamefont {J.}~\bibnamefont {Juno}},\ and\ \bibinfo
  {author} {\bibfnamefont {G.~W.}\ \bibnamefont {Hammett}},\ }\href
  {https://doi.org/10.1017/S0022377820000586} {\bibfield  {journal} {\bibinfo
  {journal} {J. Plasma Phys.}\ }\textbf {\bibinfo {volume} {86}},\ \bibinfo
  {pages} {905860403} (\bibinfo {year} {2020})}\BibitemShut {NoStop}%
\bibitem [{\citenamefont {Ji}\ and\ \citenamefont {Held}(2006)}]{Ji2006}%
  \BibitemOpen
  \bibfield  {author} {\bibinfo {author} {\bibfnamefont {J.-Y.}\ \bibnamefont
  {Ji}}\ and\ \bibinfo {author} {\bibfnamefont {E.~D.}\ \bibnamefont {Held}},\
  }\href {https://doi.org/10.1063/1.2356320} {\bibfield  {journal} {\bibinfo
  {journal} {Phys. Plasmas}\ }\textbf {\bibinfo {volume} {13}},\ \bibinfo
  {pages} {102103} (\bibinfo {year} {2006})}\BibitemShut {NoStop}%
\bibitem [{\citenamefont {Ji}\ and\ \citenamefont {Held}(2008)}]{Ji2008}%
  \BibitemOpen
  \bibfield  {author} {\bibinfo {author} {\bibfnamefont {J.-Y.}\ \bibnamefont
  {Ji}}\ and\ \bibinfo {author} {\bibfnamefont {E.~D.}\ \bibnamefont {Held}},\
  }\href {https://doi.org/10.1063/1.2977983} {\bibfield  {journal} {\bibinfo
  {journal} {Phys. Plasmas}\ }\textbf {\bibinfo {volume} {15}},\ \bibinfo
  {pages} {102101} (\bibinfo {year} {2008})}\BibitemShut {NoStop}%
\bibitem [{\citenamefont {Ji}\ and\ \citenamefont {Held}(2009)}]{Ji2009}%
  \BibitemOpen
  \bibfield  {author} {\bibinfo {author} {\bibfnamefont {J.-Y.}\ \bibnamefont
  {Ji}}\ and\ \bibinfo {author} {\bibfnamefont {E.~D.}\ \bibnamefont {Held}},\
  }\href {https://doi.org/10.1063/1.3234253} {\bibfield  {journal} {\bibinfo
  {journal} {Phys. Plasmas}\ }\textbf {\bibinfo {volume} {16}},\ \bibinfo
  {pages} {102108} (\bibinfo {year} {2009})}\BibitemShut {NoStop}%
\bibitem [{\citenamefont {Abramowitz}\ \emph {et~al.}(1965)\citenamefont
  {Abramowitz}, \citenamefont {Stegun},\ and\ \citenamefont
  {Miller}}]{Abramowitz1972}%
  \BibitemOpen
  \bibfield  {author} {\bibinfo {author} {\bibfnamefont {M.}~\bibnamefont
  {Abramowitz}}, \bibinfo {author} {\bibfnamefont {I.}~\bibnamefont {Stegun}},\
  and\ \bibinfo {author} {\bibfnamefont {D.}~\bibnamefont {Miller}},\ }\href
  {https://doi.org/10.1115/1.3625776} {\emph {\bibinfo {title} {{Handbook of
  Mathematical Functions With Formulas, Graphs and Mathematical Tables}}}}\
  (\bibinfo  {publisher} {Dover Publications, Inc},\ \bibinfo {address} {New
  York},\ \bibinfo {year} {1965})\BibitemShut {NoStop}%
\bibitem [{\citenamefont {Zocco}\ and\ \citenamefont
  {Schekochihin}(2011)}]{Zocco2011}%
  \BibitemOpen
  \bibfield  {author} {\bibinfo {author} {\bibfnamefont {A.}~\bibnamefont
  {Zocco}}\ and\ \bibinfo {author} {\bibfnamefont {A.~A.}\ \bibnamefont
  {Schekochihin}},\ }\href {https://doi.org/10.1063/1.3628639} {\bibfield
  {journal} {\bibinfo  {journal} {Phys. Plasmas}\ }\textbf {\bibinfo {volume}
  {18}},\ \bibinfo {pages} {102309} (\bibinfo {year} {2011})}\BibitemShut
  {NoStop}%
\bibitem [{\citenamefont {Loureiro}\ \emph {et~al.}(2016)\citenamefont
  {Loureiro}, \citenamefont {Dorland}, \citenamefont {Fazendeiro},
  \citenamefont {Kanekar}, \citenamefont {Mallet}, \citenamefont {Vilelas},\
  and\ \citenamefont {Zocco}}]{Loureiro2015}%
  \BibitemOpen
  \bibfield  {author} {\bibinfo {author} {\bibfnamefont {N.~F.}\ \bibnamefont
  {Loureiro}}, \bibinfo {author} {\bibfnamefont {W.}~\bibnamefont {Dorland}},
  \bibinfo {author} {\bibfnamefont {L.}~\bibnamefont {Fazendeiro}}, \bibinfo
  {author} {\bibfnamefont {A.}~\bibnamefont {Kanekar}}, \bibinfo {author}
  {\bibfnamefont {A.}~\bibnamefont {Mallet}}, \bibinfo {author} {\bibfnamefont
  {M.~S.}\ \bibnamefont {Vilelas}},\ and\ \bibinfo {author} {\bibfnamefont
  {A.}~\bibnamefont {Zocco}},\ }\href
  {https://doi.org/10.1016/j.cpc.2016.05.004} {\bibfield  {journal} {\bibinfo
  {journal} {Comput. Phys. Commun.}\ }\textbf {\bibinfo {volume} {206}},\
  \bibinfo {pages} {45} (\bibinfo {year} {2016})}\BibitemShut {NoStop}%
\bibitem [{\citenamefont {Dougherty}(1964)}]{Dougherty1964}%
  \BibitemOpen
  \bibfield  {author} {\bibinfo {author} {\bibfnamefont {J.~P.}\ \bibnamefont
  {Dougherty}},\ }\href {https://doi.org/10.1063/1.2746779} {\bibfield
  {journal} {\bibinfo  {journal} {Phys. Fluids}\ }\textbf {\bibinfo {volume}
  {7}},\ \bibinfo {pages} {1788} (\bibinfo {year} {1964})}\BibitemShut
  {NoStop}%
\bibitem [{\citenamefont {Gillis}\ and\ \citenamefont
  {Shimshoni}(1962)}]{Gillis1962}%
  \BibitemOpen
  \bibfield  {author} {\bibinfo {author} {\bibfnamefont {J.}~\bibnamefont
  {Gillis}}\ and\ \bibinfo {author} {\bibfnamefont {M.}~\bibnamefont
  {Shimshoni}},\ }\href {https://doi.org/10.2307/2003810} {\bibfield  {journal}
  {\bibinfo  {journal} {Math. Comput.}\ }\textbf {\bibinfo {volume} {16}},\
  \bibinfo {pages} {50} (\bibinfo {year} {1962})}\BibitemShut {NoStop}%
\bibitem [{\citenamefont {Askey}\ and\ \citenamefont
  {Gasper}(1977)}]{Askey1977}%
  \BibitemOpen
  \bibfield  {author} {\bibinfo {author} {\bibfnamefont {R.}~\bibnamefont
  {Askey}}\ and\ \bibinfo {author} {\bibfnamefont {G.}~\bibnamefont {Gasper}},\
  }\href {https://doi.org/10.1007/BF02813297} {\bibfield  {journal} {\bibinfo
  {journal} {J. d'Analyse Math{\'{e}}matique}\ }\textbf {\bibinfo {volume}
  {31}},\ \bibinfo {pages} {48} (\bibinfo {year} {1977})}\BibitemShut {NoStop}%
\bibitem [{\citenamefont {Kleindienst}\ and\ \citenamefont
  {Luchow}(1993)}]{Kleindienst1993}%
  \BibitemOpen
  \bibfield  {author} {\bibinfo {author} {\bibfnamefont {H.}~\bibnamefont
  {Kleindienst}}\ and\ \bibinfo {author} {\bibfnamefont {A.}~\bibnamefont
  {Luchow}},\ }\href {https://doi.org/10.1002/qua.560480405} {\bibfield
  {journal} {\bibinfo  {journal} {Int. J. Quantum Chem.}\ }\textbf {\bibinfo
  {volume} {48}},\ \bibinfo {pages} {239} (\bibinfo {year} {1993})}\BibitemShut
  {NoStop}%
\bibitem [{\citenamefont {Khabibrakhmanov}\ and\ \citenamefont
  {Summers}(1998)}]{Khabibrakhmanov1998}%
  \BibitemOpen
  \bibfield  {author} {\bibinfo {author} {\bibfnamefont {I.~K.}\ \bibnamefont
  {Khabibrakhmanov}}\ and\ \bibinfo {author} {\bibfnamefont {D.}~\bibnamefont
  {Summers}},\ }\href {https://doi.org/10.1016/S0898-1221(98)00117-5}
  {\bibfield  {journal} {\bibinfo  {journal} {Comput. Math. with Appl.}\
  }\textbf {\bibinfo {volume} {36}},\ \bibinfo {pages} {65} (\bibinfo {year}
  {1998})}\BibitemShut {NoStop}%
\bibitem [{\citenamefont {Erdelyi}(1936)}]{Erdelyi1936}%
  \BibitemOpen
  \bibfield  {author} {\bibinfo {author} {\bibfnamefont {A.}~\bibnamefont
  {Erdelyi}},\ }\href {https://doi.org/10.1007/BF01218891} {\bibfield
  {journal} {\bibinfo  {journal} {Math. Zeitschrift}\ }\textbf {\bibinfo
  {volume} {40}},\ \bibinfo {pages} {693} (\bibinfo {year} {1936})}\BibitemShut
  {NoStop}%
\bibitem [{\citenamefont {Gordon}(1929)}]{Gordon1929}%
  \BibitemOpen
  \bibfield  {author} {\bibinfo {author} {\bibfnamefont {W.}~\bibnamefont
  {Gordon}},\ }\href {https://doi.org/10.1002/andp.19293940807} {\bibfield
  {journal} {\bibinfo  {journal} {Ann. Phys.}\ }\textbf {\bibinfo {volume}
  {394}},\ \bibinfo {pages} {1031} (\bibinfo {year} {1929})}\BibitemShut
  {NoStop}%
\end{thebibliography}%

\end{document}